\documentclass[a4paper,11pt]{article}
\pdfoutput=1 

\usepackage{jheppub} 

\usepackage{
	amsmath,
	mathtools,
	physics,
	subcaption,
	tensor
}
\usepackage[dvipsnames]{xcolor}

\newcommand{\ee}{\ensuremath{\mathrm{e}}}
\renewcommand{\imath}{\mathrm{i}}

\newcommand{\HypergeometricPFQ}[5]{\ensuremath{{}_{#1}\hskip-1ptF_{#2}\left(#3;#4;#5	\right)}}
\newcommand{\HypergeometricZeroFOne}[2]{\ensuremath{{}_{0}\hskip-.5ptF_{1}\left(#1;#2	\right)}}
\newcommand{\HypergeometricPFQtilde}[5]{\ensuremath{{}_{#1}\hskip-1pt\tilde{F}_{#2}\left(#3;#4;#5	\right)}}
\newcommand{\HypergeometricPFQDerivative}[6]{\ensuremath{{}_{#2}^{\phantom{0}}\hskip-1pt F_{#3}^{#1}\left(#4;#5;#6	\right)}}

\newcommand{\dimd}{\ensuremath{d}}
\newcommand{\GN}{\ensuremath{G_N}}
\newcommand{\CC}{\ensuremath{\Lambda}}
\newcommand{\aRR}{\ensuremath{\alpha_{R}}}
\newcommand{\aCC}{\ensuremath{\alpha_{C}}}
\newcommand{\aE}{\ensuremath{\alpha_{E}}}
\newcommand{\aR}[1]{\ensuremath{\alpha_{#1}}}
\newcommand{\alphagf}{\ensuremath{\alpha_{\text{gf}}}}
\newcommand{\betagf}{\ensuremath{\beta_{\text{gf}}}}

\newcommand{\proper}[1]{\ensuremath{\mathfrak{#1}}}

\newcommand{\deS}{de~Sitter}
\newcommand{\DeS}{De~Sitter}
\newcommand{\QuadG}{QuadG}

\title{Quadratic gravity potentials in de Sitter spacetime from Feynman diagrams}

\author{Renata Ferrero}
\author{and Chris Ripken}

\affiliation{Institute of Physics (THEP), University of Mainz,
	\\Staudingerweg 7, D-55128 Mainz, Germany\\ \\MITP-22-104}

\emailAdd{rferrero@uni-mainz.de}

\abstract{We employ a manifestly covariant formalism to compute the tree-level amputated Green's function of non-minimally coupled scalar fields in quadratic gravity in a de Sitter background. We study this Green's function in the adiabatic limit, and construct the classical Newtonian potential. At short distances, the flat-spacetime Yukawa potential is reproduced, while the curvature gives rise to corrections to the potential at large distances. Beyond the Hubble radius, the potential vanishes identically, in agreement with the causal structure of de Sitter spacetime. For sub-Hubble distances, we investigate whether the modifications to the potential reproduce Modified Newtonian Dynamics.}

\begin{document} 
	\maketitle
	\flushbottom
	
	\section{Introduction}
	\DeS{} spacetime plays a vital role in our current understanding of the Universe at different epochs. In early-time cosmology, \deS{} spacetime provides an accurate description of an inflationary universe, with evidence provided by the Cosmic Microwave Background (CMB) \cite{2020}. Also at late cosmic times, observations of distant supernovae suggest an accelerated expansion of the universe, modeled by a \deS{} spacetime \cite{SupernovaCosmologyProject:1998vns,SupernovaSearchTeam:1998fmf}.
	
	On the other hand, there are still significant open questions regarding \deS{} spacetime. First, we know that it arises in General Relativity (GR) as the solution to Einstein's equation including a cosmological constant $\Lambda$. Usually attributed to so-called ``dark energy'' or an intrinsic ``vacuum energy'', the origin of the cosmological constant, and its small present-day value of $67.66$ (km/s)/Mpc $=10^{-122}\;\ell_\text{P}^2$ \cite{Barrow:2011zp} remains a mystery \cite{Martin:2012bt}.
	
	How to reconcile GR with quantum theory remains is currently unknown. In a flat background, perturbative quantization leads to a perturbatively non-renormalizable quantum theory \cite{tHooft:1974toh,Goroff:1985sz, Goroff:1985th}. There are, by now, many ideas  on how to resolve this, and construct a theory of Quantum Gravity. Of these approaches, string theory \cite{polchinski_1998, Tong:2009np}, loop quantum gravity \cite{Rovelli:1997yv, Dittrich:2004bn}, and asymptotically safe quantum gravity (see for instance \cite{Wetterich:1992yh, Reuter:1996cp, Reuter:2019byg, Percacci:2017fkn, Eichhorn:2018yfc, Pawlowski:2020qer, Ferrero:2022hor} for its continuum  and \cite{Ambjorn:2012jv,Loll:2019rdj} for its discrete implementation) are notable examples. However, no candidate has been undisputedly successful so far.
	
Green's functions in curved spacetime may provide insight into how to construct a consistent theory of quantum gravity. 
Being at the basis of observables in Quantum Field Theory, a proper understanding of scattering amplitudes in a gravitating background may help uncover how to incorporate long-distance curvature effects into quantum theory.
Along another direction, by treating gravity as an effective field theory, scattering amplitudes involving loops introduce classical (post-Newtonian) corrections and quantum corrections to the gravitational potential \cite{Donoghue:1993eb,Hamber:1995cq,Bjerrum-Bohr:2002gqz, Carrillo-Gonzalez:2021mqj, Bjerrum-Bohr:2022blt,Kosower:2022yvp}.
	
	Green's functions in \deS{} spacetime are an important testing ground for such a program. Being the simplest non-trivially curved spacetime, amplitudes in \deS{} spacetime already exhibit the main challenges of quantizing in a generic curved spacetime, such as the absence of a unique invariant vacuum \cite{PhysRevD.32.3136,birrell_davies_1982,Mukhanov:2007zz, Akhmedov:2019cfd}, the obstructions in defining an $S$-matrix \cite{Witten:2001kn, Bousso:2004tv,Marolf:2012kh, Mandal:2019bdu,Giddings:2009gj, deRham:2022hpx} and the presence of singularities in correlation functions \cite{allen2,Allen1987THEGP, Allen1987AnEO,Floratos:1987ek, PhysRevD.16.245, PhysRevD.45.2013, Akhmedov:2011pj, Akhmedov:2013vka, Akhmedov:2019esv, Akhmedov:2020qxd,Lochan:2018pzs,Singh:2013dia, Kim:2010cb, Bros:1994dn, Bros:1995js,Bros:1998ik,Higuchi:2010xt,Higuchi:2002sc,Fukuma:2013mx,Arkani-Hamed:2015bza,Arkani-Hamed:2018kmz, Cacciatori:2007in}.
	
		Because of the absence of a unique invariant vacuum, the notion of scattering amplitude in curved spacetime is, in general, not well-defined and the construction is not free of ambiguities. On the other hand, the construction of  well-defined Green's functions in curved spacetime is possible, by picking a particular vacuum state and  following the quantum field theoretical rules. 
We expect then that these objects encode physical information about particle scattering processes.
		As a guiding principle, one could aim to reproduce the flat spacetime result in the flat spacetime limit.  This is the strategy adopted in this paper.

	In \cite{Ferrero:2021lhd}, we put forward a novel technique to compute Green's functions in a \deS{} background. Key in this approach is the observation that Feynman diagrams can be represented by differential operators. In flat spacetime, these are commuting, so that we conveniently can go to momentum space. In curved spacetime, however, the non-commuting character of differential operators requires a more careful treatment.
	
	The main result of \cite{Ferrero:2021lhd} was the tree-level amplitude amputated Green's function of the gravity-mediated scattering of two minimally coupled massive scalars in \deS{} spacetime. Here the gravitational dynamics is specified by the Einstein-Hilbert action. Employing an expansion around infinite scalar masses, we use this amplitude to compute the \deS{} spacetime generalization of the Newtonian potential. Remarkable features of this potential are curvature-dependent corrections corresponding to a repulsive force, consistent with the picture of \deS{} spacetime as an expanding universe, and a vanishing potential at super-Hubble distances, encoding a causal separation by the \deS{} horizon.
	
	In this paper, we extend this computation to quadratic gravity (\QuadG) \cite{Sexl, Havas,  Donoghue:2021cza,Salvio:2019ewf}. In the pure-gravity sector, we include all terms in the action up to four derivatives of the metric, while we also allow for a non-minimal $R\phi\phi$-coupling between the scalar fields and the Ricci scalar.
	
	The extension to \QuadG{} is interesting for a number of reasons. First, \QuadG{} successfully provides the dynamics of primordial graviton fluctuations \cite{Starobinsky:1980te, Ketov:2012jt}. Its predictions for inflationary power spectra and spectral indices are in excellent agreement with observational data \cite{Koshelev:2017tvv, Koshelev:2020foq}, and lead to constraints on the gravitational couplings \cite{Vilenkin:1985md, Berry:2011pb, Cembranos:2008gj, Kapner:2006si}.
	
	Second, \QuadG{} is attractive from a theoretical perspective. As was shown by Stelle \cite{Stelle:1976gc, Stelle:1977ry} it is perturbatively renormalizable in flat spacetime, in contrast to Einstein-Hilbert gravity. However, this does not come for free. The addition of four-derivative terms in the action implies that the resulting Hamiltonian is unbounded from below. Classically, this leads to the Ostrogradsky instability \cite{Ostrogradsky}. At the quantum level, it is signaled by the appearance of a massive spin-2 ghost \cite{Antoniadis, Tomboulis:1977jk, Donoghue_ghost}.
	
	Third, as we demonstrated in \cite{Ferrero:2021lhd}, the Newtonian potential receives corrections due to the background curvature. In GR, these corrections are suppressed by the inverse \deS{} radius. Therefore, the potential meets all observational tests on length scales ranging from table-top to solar system \cite{Sokolowski:2008kf,Capozziello:2009ss,Naf:2010zy,Conroy:2014eja,Alvarez-Gaume:2015rwa,Edholm:2016hbt,Cheung:2018wkq}. In \QuadG, however, the new coupling constants induce additional length scales, leading possibly to corrections to the potential at much smaller lengths. Thus, the computation of the Newtonian potential in \QuadG{} may lead to additional constraints on the non-minimal couplings.
	
	Replacing the classical Newtonian potential by a more general potential falls into the class of Modified Newtonian Dynamics \cite{Milgrom1983AMO,Milgrom:2011kx,Famaey:2011kh, deAlmeida:2018kwq}. This has been used to successfully account for the observed properties of rotation curves of galaxies \cite{Brouwer:2021nsr, mond1, Blanchet:2011pv}. Therefore, it is seen as an alternative to the dark matter (DM) hypothesis. In our setting, we will test whether the modifications to the Newtonian potential due to QuadG in \deS{} spacetime give rise to a DM-like MOND scenario.

	This paper is organized as follows. In \autoref{sec:Yukawa} we derive the Yukawa potential and discuss the analogous construction of the gravitational potential. In \autoref{sec:amplitudefunctional}, we describe how to compute the Green's function of a two-to-two scalar process in a \deS{} background in \QuadG. In \autoref{sec:adiabatic-amplitude}, we compute the adiabatic expansion of this Green's function. This is translated to the nonrelativistic potential in \autoref{sec:potential}. We end with a brief summary and concluding remarks in \autoref{sec:conclusion}. We collect our conventions for \deS{} spacetime in \autoref{app:conventions}. Details regarding the adiabatic expansion in general and the Green's function are relegated to \autoref{app:adiabatic} and \autoref{app:details}, respectively.
	
	\section{Warmup: the Yukawa potential}\label{sec:Yukawa}
	Before constructing the Green's function  of a graviton-mediated scalar-to-scalar process, we will set the stage by following the standard derivation of the the Yukawa potential for a massive $\phi^3$-scalar theory  scalar-to-scalar scattering process in flat spacetime.
	
	The Yukawa potential can be derived as the lowest order amplitude of the interaction of a pair of scalars. The Yukawa interaction couples them to another exchanged scalar field with the interaction term $g \; \phi^3 $. The action of the three scalars then  reads
	\begin{equation}
		S_\text{} = \int \text{d}^d x \;\left(\frac{1}{2}\partial_\mu \phi \partial^\mu \phi +\frac{1}{2}\partial_\mu \chi \partial^\mu \chi  - \frac{m^2}{2}\phi^2  - \frac{M^2}{2}\chi^2-g\; \phi^2 \chi \right)
		\,\text{.}
	\end{equation}
	
	\begin{figure}
		\centering
		\includegraphics[scale = 1]{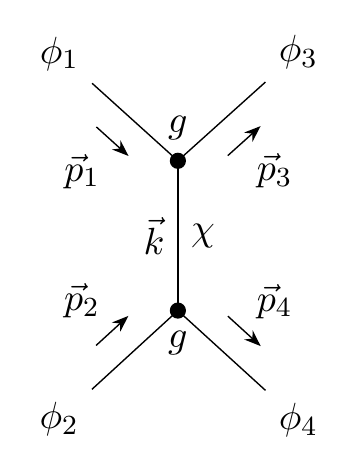}
		\caption{Tree-level scalar-to-scalar $\phi^2$-scattering amplitude for the process $\phi_1\phi_2 \to \phi_3 \phi_4$ ($t$-channel). Time flows from the left to the right.}
		\label{fig:feynmandiagramYukawa}
	\end{figure}
	
	The scattering amplitude for two scalars, one with initial momentum $p_1$ and the other with momentum 
	$p_2$, exchanging a scalar $\chi$ with momentum $k$ (see \autoref{fig:feynmandiagramYukawa}), is constructed by applying the Feynman rules: 
	\begin{enumerate}
		\item For each vertex associate a factor of $g$ with the amplitude; since this diagram has two vertices, the total amplitude will have a factor of $g^2$;
		\item The Feynman rule for a particle exchange is to use the propagator; the propagator for a  scalar  with mass $M$ is $
		-\frac{	4\pi}{k^2+M^2} $;
		\item Thus, we see that the Feynman amplitude for this interaction is nothing more than
		\begin{equation}
			\tilde{V}(k)=	-g^2\frac{	4\pi}{k^2+M^2} \stackrel{M\gg |\vec{k}|}{=}-g^2\frac{	4\pi}{|\vec k|^2+M^2}
		\end{equation}
		where in the last equality we have applied the so-called \textit{Born approximation}: we assumed that the mass is much greater than the exchanged 3-momentum.
	\end{enumerate}
	This is seen to be the Fourier transform of the Yukawa potential by examining its Fourier transform:
	\begin{equation}
		V(r) = \frac{g^2}{(2\pi)^3} \int e^{\imath\vec{k} \cdot \vec{r}}\frac{	4\pi}{|\vec k|^2+M^2} \text{d} ^3 \vec k = -g^2\; \frac{e^{-M \;r}}{r}
		\,\text{.}
	\end{equation}
	The potential is monotonically increasing in $r$ and it is negative, implying that the force is attractive.
	
	\subsection{Gravitational potential around flat spacetime}
	In the following subsection we will extend this computation to gravitational interactions \cite{Bjerrum-Bohr:2002gqz} by sketching the derivation of the Newtonian potential, and to the Quadratic Gravity action. Here, both cases are treated in flat spacetime.
	
Ref. \cite{Bjerrum-Bohr:2002gqz} sketched how the Newtonian $1/r$-potential could be derived by performing the Born approximation to the following scattering process: they considered a two-to-two-scalars graviton mediated scattering amplitude. Due to the masslessness of the exchanged graviton, this resulted in a $1/r$ interaction potential.

 Analogously, by studying the propagating modes corresponding to Quadratic Gravity around flat spacetime, the same Yukawa terms should arise \cite{Alvarez-Gaume:2015rwa}. Consider the four-derivative gravitational action (in the metric $\hat g_{\mu \nu}$) in 4 dimensions and two non-minimally coupled scalar fields
	\begin{equation} 
		S[\hat g, \phi, \chi]	=	\frac{1}{16\pi G_N}	\int \dd[d]{x}	\sqrt{-\hat g}	\,	\bigg(\begin{aligned}[t]&	-2\CC + \hat{R}	+		\frac{\aRR}{6}	\hat{R}^2	
			-		\frac{\aCC}{2}	\hat{C}_{\mu\nu\rho\sigma}\hat{C}^{\mu\nu\rho\sigma}		\\&-\frac{1}{2}		\phi	\left(	\hat\square + m^2_\phi+	\aR{\phi}	\hat{R}	\right)	\phi -\frac{1}{2}		\chi	\left(	\hat\square + m^2_\chi+	\aR{\chi}	\hat{R}	\right)	\chi\bigg)
			\,\text{.}
		\end{aligned}
	\end{equation}
The flat-spacetime amplitude of a $\phi\chi\to \phi \chi$ graviton-mediated scattering amplitude (see \autoref{fig:diagram}) can be computed applying the recipe of the previous subsection. By expanding  quadratically the action  around Minkowski space $\hat g_{\mu \nu} =\eta_{\mu \nu}+ h_{\mu \nu}$  and choosing the de~Donder gauge fixing one can derive the different contributions for the propagators. They can be interpreted as the propagator of  a massless spin-0 and spin-2 particle (the graviton) and a  spin-0 particle with mass parameter $M^2 = \alpha_R^{-1}$ and a  spin-2 particle with mass $M^2=\alpha_C^{-1}$. These appear in the amplitude with suitable prefactors and signs. Schematically, denoting $q$ the momentum of the exchanged particle, the different contributions to the  propagators are:
	\begin{align}
	\text{spin-2:}	\qquad&\frac{1}{q^2}, \qquad\frac{1}{q^2 + \alpha_C^{-1}}
	\\
	\text{spin-0:}	\qquad&\frac{1}{q^2}, \qquad\frac{1}{q^2 + \alpha_R^{-1}}
\end{align}

	The related potential can be obtained by performing the Fourier transform of the amplitude. We obtain
	\begin{equation}\
		V_{\text{Mink}}(r) \propto G_N  \left(	-\frac{1}{r}	- \frac{1}{3} \frac{1}{r}\ee^{-r/\sqrt{\aRR}} + \frac{4}{3}	\frac{1}{r}\ee^{-r/\sqrt{\aCC}}	\right)\;.
	\end{equation}
	The first term represents the Newtonian contribution, while the other two terms are Yukawa potentials due to the presence of quadratic interactions. 
	
It is important to emphasize, that the two additional Yukawa contributions appear with two opposite signs. The $\aCC$ term has a positive sign, meaning that it competes with the Newtonian contribution and is repulsive. Demanding that the potential should reproduce the Newtonian behavior at short distances will impose  constraints on the value of $\aCC$.

	\section{The quadratic gravity Green's function in de Sitter spacetime}\label{sec:amplitudefunctional}
\begin{figure}\centering
		\includegraphics[scale=1]{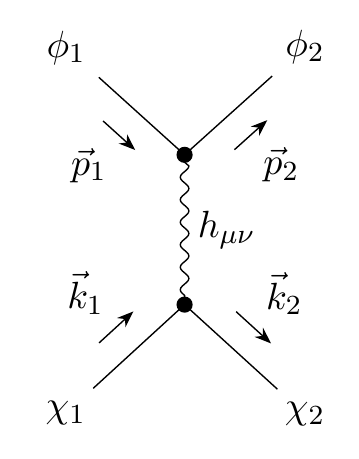}
		\caption{Feynman diagram of the scattering process $\phi\chi\to\phi\chi$ mediated by a graviton $h_{\mu\nu}$.}
		\label{fig:diagram}
	\end{figure}

	In this section, we compute the Green's function of a two-to-two particle scattering process in \deS{} spacetime. We consider the graviton-mediated scattering of two scalars $\phi$ and $\chi$, in the process $\phi \chi \to \phi \chi$. At tree level, this process is described by the $t$-channel Feynman diagram, depicted in \autoref{fig:diagram}. Here we generalize this Feynman diagram to \deS{} spacetime.

	We first give a heuristic interpretation of Green's functions in curved spacetime in \autoref{sec:gedankenexperiment}. We then set the stage by specifying the action in \autoref{sec:action}. We provide details on how to compute the   Green's function in \deS{} spacetime in \autoref{sec:amplitudefunctionalcomputation}. The resulting  amputated Green's function is then presented in \autoref{sec:amplitudefunctionalresult}. We finish this section with a few remarks in \autoref{sec:amplitudefunctionaldiscussion}.
	
	\subsection{Gedankenexperiment}\label{sec:gedankenexperiment}
		In this section, we will briefly discuss the interpretation of scattering processes from an experimental point of view. Scattering amplitudes represent a technical tool developed with the purpose to connect QFT and collider experiments. However, as previously mentioned, due to the lack of a description of an $S$-matrix in QFT in curved spacetime, the generalization to possible collision experiments in general spacetimes has not been performed yet.
		
		By constructing  Green's functions in de Sitter spacetime in a fully covariant way, we can mimic the connection to experiments generally used in Minkowski spacetime. In this way we circumvent the lack of $S$-matrix elements in de Sitter. For the sake of simplicity, here we will restrict the discussion to GR interactions, considering in the gravitational action only the Einstein–Hilbert term (neglecting higher curvature interactions).
		
	The resulting picture is the following: we prepare  two heavy-mass particles in a de Sitter space and localize them in space. The two particles scatter. Once the process has finished, we detect where the outgoing particles have ended up.
	
	In particular we will work in the expanding Poincaré patch. Here the particles will be  Bunch–Davies waves, i.e.,  the wave functions solve the wave equation in the expanding Poincaré patch. The Bunch–Davies vacuum  is a state that has no positive energy excitations at the past infinity of the expanding Poincaré patch. In fact near the boundary we can define what we mean by particle and what we mean by positive energy, because every momentum experiences infinite blue shift towards the past of the patch. Here, high energy harmonics are not sensitive to the comparatively small curvature of the background space and behave as if they are in flat space.
	
	Due to the spatial homogeneity of the conformally flat patch and also of the initial states that we consider, it will be natural to perform the Fourier transformation along the spatial directions. The explicit computation of the  Green's function will allow us to pass from momentum space to position space and to construct the  potential.

	\subsection{Action}\label{sec:action}
	We start our computation by introducing the action $S$. We use an action that is a functional of the (dynamical) spacetime metric $\hat g$ and two scalar fields $\phi$ and $\chi$. In order to consistently handle the gauge degrees of freedom in the metric fluctuations, we employ the background field method, depending on the background metric $\bar g$. We denote quantities defined with respect to the dynamical metric $\hat g$ by a hat, while objects defined using the background metric $\bar g$ are denoted by a bar. The action is then of the following form:
	\begin{equation}\label{eq:S}
		S	=	S_{\text{grav}}[\hat g]	+	 S_{\text{gf}}[\hat g;\bar g]	+	S_{\text{sc}}[\phi, \hat g]	+	S_{\text{sc}}[\chi,\hat g]
		\,\text{.}
	\end{equation}
	Let us now  explicitly formulate each contribution to the action. For the gravitational interaction $S_{\text{grav}}$, we take the four-derivative action
	\begin{equation} \label{eq:Sgrav}
		S_{\text{grav}}[\hat g]	=	\frac{1}{16\pi \GN}	\int \dd[\dimd]{x}	\sqrt{-\hat g}	\,	\bigg(\begin{aligned}[t]&	-2\CC + \hat{R}	+	\aRR	\frac{\dimd-2}{4(\dimd-1)}	\hat{R}^2	
			\\&
			-	\aCC	\frac{\dimd-2}{4(\dimd-3)}	\hat{C}_{\mu\nu\rho\sigma}\hat{C}^{\mu\nu\rho\sigma}	+	\frac{1}{2(\dimd-3)(\dimd-4)}\aE E_4	\bigg)
			\,\text{.}
		\end{aligned}
	\end{equation}
	For the time being, no specification of the sign of the cosmological constant $\Lambda$ is required.
	
	In \eqref{eq:Sgrav}, $E_4$ is the Euler density in four dimensions:
	\begin{equation}
		E_4	=	\hat{R}^2	-	4 \hat{R}_{\mu\nu} \hat{R}^{\mu\nu}	+	\hat{R}_{\alpha\beta\gamma\delta} \hat{R}^{\alpha\beta\gamma\delta}
		\,\text{.}
	\end{equation}
	For $\dimd=4$, the integral over $E_4$ is a topological invariant. Thus, for this dimension we set $\aE$ to zero, since it will not contribute to any dynamical quantity. In a similar fashion, the $C^2$ and $R^2$ are non-dynamical for $\dimd\leq 3 $ and $\dimd\leq 2$ dimensions, respectively. We therefore set the $\aCC$ and $\aRR$ couplings to zero in these dimensions, too.
	In flat spacetime, the $\aRR$ and $\aCC$ couplings induce massive spin-0 and spin-2 poles in the graviton propagator. We have chosen the normalization in such a way that in flat spacetime, $\aRR$ denotes the inverse mass of the spin-0 pole, while $\aCC$ denotes the inverse mass of the spin-2 pole \cite{Stelle:1976gc}.
	
	Next, we specify in \eqref{eq:S} a de~Donder type gauge fixing action, necessary to obtain a well-defined graviton propagator:
	\begin{equation}
		S_{\text{gf}}[\hat g;\bar g] = -\frac{1}{16\pi \GN}\frac{1}{2\alphagf}	\int \dd[\dimd]{x} \sqrt{- \bar g}	\, \bar g^{\mu\nu}	\mathcal F_\mu[\hat g, \bar g]	\mathcal F_\nu[\hat g,\bar g]
		\,\text{,}
	\end{equation}
	where  the gauge fixing operator is defined as
	\begin{equation}
		\mathcal F_\mu[\hat g,\bar g]	=	\delta_\mu^\beta	\bar g^{\rho\sigma}	\bar \nabla_\rho	\hat g_{\sigma \beta}	-	\frac{1+\betagf}{\dimd}  \bar g^{\alpha\beta}	\bar \nabla_\mu \hat 	g_{\alpha\beta}
		\,\text{.}
	\end{equation}
	Here the gauge fixing parameters $\alphagf$ and $\betagf$ generalize the de~Donder gauge, and allow to track gauge dependence explicitly.
	The Faddeev-Popov ghosts will not contribute to the  Green's function, for the reason that they do not couple to the scalar fields. The specific form of their action is therefore not needed for our purposes.
	
	Finally, the matter sector  is given by non-minimally coupled massive scalar fields $\phi$ and $\chi$, whose action takes the form
	\begin{equation}
		S_{\text{sc}}[\phi, \hat g]	=	-\frac{1}{2}	\int \dd[\dimd]{x}	\sqrt{-\hat g}	\,	\phi	\left(	\hat\square + m_\phi^2	+	\aR{\phi}	\hat{R}	\right)	\phi
		\,\text{.}
	\end{equation}
	Here we have expressed the covariant d'Alembertian by $\hat \square = - \hat g^{\mu\nu} \hat\nabla_\mu \hat\nabla_\nu$.

	\subsection{Computation of the Green's functional} \label{sec:amplitudefunctionalcomputation}
	We will now give a brief description of how to compute the Feynman diagram in a curved background geometry. We follow the computation from \cite{Ferrero:2021lhd}; we refer to this paper for additional details.
	
	The tree-level Green's function associated to the diagram in \autoref{fig:diagram} is obtained by contracting the graviton-scalar-scalar 3-point vertices with the graviton propagator.
	These are computed by taking functional derivatives of the action $S$ with respect to the fields, and projecting onto a solution to the equation of motion. For the scalar fields, the equation of motion is given by the (non-minimal) Klein-Gordon equation:
	\begin{align}\label{eq:kgequation}
		\hat\square \phi = -(m_\phi^2 + \aR{\phi} \hat R) \phi
		\,\text{,}&\qquad&
		\hat\square \chi = -(m_\chi^2 + \aR{\chi} \hat R) \chi
		\,\text{.}
	\end{align}
	We observe that $\phi = \chi =0 $ is a solution to the equation of motion. The equation of motion for the metric is that of \QuadG. We will look for \deS{} solutions, which means that the constant Ricci scalar is the only nonzero curvature tensor. The Ricci scalar $R$ is then given by the quadratic equation
	\begin{equation}\label{eq:desittereom}
		R = \frac{2\dimd}{\dimd-2}\Lambda -	\frac{\dimd-4}{4(\dimd-1)}	\aRR R^2
		\,\text{.}
	\end{equation}
	We  write
	\begin{equation}
		\hat g_{\mu\nu}	=	\bar g_{\mu\nu} + h_{\mu\nu}
		\,\text{,}
	\end{equation}
	where $\bar g_{\mu\nu}$ is a \deS{} metric whose Ricci scalar $\bar R$ satisfies \eqref{eq:desittereom}. Functional derivatives with respect to $\hat g$ are then conveniently computed by expanding around $\bar g_{\mu\nu}$.
	
	By taking one functional derivative with respect to the metric 	and  two with respect to the scalar fields,   we obtain the three-point vertices. Subsequently, we set all fields to their background values. Thus, we define
	\begin{align}\label{eq:3pf}
		\mathcal{T}^{(h\phi\phi)}
		=
		\eval{\frac{\delta^3 S}{\delta h \delta \phi \delta \phi}	}_{\substack{\hat g=\bar g \\\phi=\chi=0}}
		\,\text{,}&\qquad&
		\mathcal{T}^{(h\chi\chi)}
		=
		\eval{\frac{\delta^3 S}{\delta h \delta \chi \delta \chi}	}_{\substack{\hat g=\bar g \\\phi=\chi=0}}
		\,\text{.}
	\end{align}
	The graviton propagator is given by the inverse of the two-point function:
	\begin{equation}
		\mathcal{G}^{(hh)}
		=
		\eval{\left[	\frac{\delta^2 S}{\delta h\delta h}	\right]^{-1}	}_{\substack{\hat g=\bar g \\\phi=\chi=0}}
		\,\text{.}
	\end{equation}
	Note that in order to make the inversion well-defined, it was necessary to include the gauge fixing action $S_{\text{gf}}$ in \eqref{eq:S}.
	
	The tree-level Green's function associated to the diagram in \autoref{fig:diagram} is now obtained by contracting the three-point vertices with the graviton propagator. Rather than treating the Green's function as a multilinear operator acting on fields, it is convenient to contract the Green's function  with four test fields:
	\begin{equation}\label{eq:amplitudefunctionaldefinition}
		\mathcal{A}[\chi_1,\chi_2,\phi_1,\phi_2]
		=
		\int \dd[\dimd]{x}\sqrt{-\bar g}	\,	\left[ \mathcal{T}^{(h\chi\chi)}\left(\chi_1,\chi_2\right) \right]^{\mu\nu}	\tensor{\left[\mathcal G^{(hh)} \right]}{_{\mu\nu}^{\rho\sigma}}	\left[ \mathcal{T}^{(h\phi\phi)}\left(\phi_1,\phi_2\right) \right]_{\rho\sigma}
		\,\text{.}
	\end{equation}
	Here the $\mathcal{T}^{(h\chi\chi)}(\chi_1, \chi_2)$ denotes the three-point vertex  \eqref{eq:3pf} acting on two test fields $\chi_1$ and $\chi_2$. The object $	\mathcal{A}$ will be referred to as Green's functional. Since \eqref{eq:amplitudefunctionaldefinition} only contains objects in the \deS{} background, to  simplify the notation, we will omit the bar to denote the background metric and its derived objects.
	
	At this point, we note that the building blocks of \eqref{eq:amplitudefunctionaldefinition} are noncommuting linear differential operators acting on the scalar fields and metric fluctuations. This can be contrasted to two common methods used to compute Feynman diagrams. First of all, in flat spacetime, one has access to momentum-space techniques to easily convert the vertices and propagators to a number. In curved spacetime, derivatives do not commute and a description in terms of Fourier transforms does not exist, in general. Hence, noncommuting differential operators are the natural generalization of flat-spacetime momenta. Second, considering vertices and propagators as operators greatly simplifies composition and contraction of Feynman diagram elements. The differential operator language has the advantage over computations using position-space integral kernels like Green's functions, which would imply performing lengthy integrals over \deS{} spacetime. 
	
	We computed the Green's functional \eqref{eq:amplitudefunctionaldefinition} using the Mathematica tensor algebra package \emph{xAct} \cite{Brizuela:2008ra,DBLP:journals/corr/abs-0803-0862,Nutma:2013zea,xActwebpage}. We used  the procedure outlined in \cite{Ferrero:2021lhd} to construct the graviton propagator and to sort and contract the covariant derivatives.
	
	The Green's functional \eqref{eq:amplitudefunctionaldefinition} consists of covariant derivatives acting on the scalar fields, and scalar curvatures. The Green's function can be simplified using the on-shell conditions \eqref{eq:kgequation}, integration by parts and the commutation techniques developed in \cite{Knorr:2020bjm,Ferrero:2021lhd}.
	These methods can be used to bring the Green's functional to a standard form.	We will require that the Green's functional is manifestly symmetric in the external fields. Since the vertices contain at most two uncontracted derivatives, we infer that the Green's function can be built using a scalar vertex structure, and a symmetric rank-two tensor  vertex structure. For the scalar vertex, it is convenient to define
	\begin{equation}\label{eq:scalarvertex}
		V[\phi_1,\phi_2]
		=
		\left[	m_\phi^2	+	(\dimd-1)\left(\aR{\phi}-\frac{\dimd-2}{4(\dimd-1)}\right)\square	\right]	\phi_1	\phi_2
		\,\text{,}
	\end{equation}
	where the d'Alembertian acts on the product $\phi_1\phi_2$. For the tensor vertex we choose
	\begin{equation}\label{eq:tensorvertex}
		T_{\mu\nu}[\phi_1,\phi_2]
		=
		(\nabla_{(\mu}\phi_1)(\nabla_{\nu)}\phi_2)	-	\frac{1}{\dimd}	g_{\mu\nu}(\nabla_\gamma \phi_1)(\nabla^\gamma \phi_2)
		-	\aR{\phi}	\left[	\nabla_\mu\nabla_\nu	+	\frac{1}{\dimd}	g_{\mu\nu}\square\right]\phi_1\phi_2
		\,\text{.}
	\end{equation}
	
	Finally, it will be convenient to define the operators
	\begin{equation}
		\mathcal{G}_r(\square;\zeta)
		=
		\Big(	\square	+	\big(r(r+\dimd-2) + \zeta	\big)H^2	\Big)^{-1}
		\,\text{.}
	\end{equation}
	Here we define the Hubble parameter by $H^2=\frac{R}{\dimd(\dimd-1)}$, following the conventions from \autoref{app:conventions}.
	We will refer to the operator $\mathcal{G}_r(\square;\zeta)$ as propagator. The label $r$ denotes the rank of the tensor it acts upon. We will refer to the dimensionless number $\zeta$ as the mass parameter; using slightly sloppy terminology, it parameterizes the graviton mass and can be related to the unitary irreducible representations of the graviton degrees of freedom \cite{Garidi:2003ys,Pejhan:2018ofn, Joung:2006gj, Joung:2007je, Sengor:2019mbz, Gazeau:2006gq}.
	
	\subsection{Result}\label{sec:amplitudefunctionalresult}
	We will now present the resulting Green's functional. This constitutes the first major result of this paper.	The Green's functional \eqref{eq:amplitudefunctionaldefinition} is given by
	\begin{equation}
		\begin{aligned}\mathcal{A} = 16\pi \GN \, c(\aRR,\aE)	\bigg[ &
			-	\frac{\zeta_C+\dimd}{\zeta_C-\zeta_h}	\Big(	\mathcal{A}_2(\zeta_h)	-	\mathcal{A}_2(\zeta_C)	\Big)
			\\&
			+	\frac{1}{\dimd(\dimd-1)}	\frac{\zeta_C}{\zeta_C-\zeta_h}	\Big( \mathcal{A}_0(\zeta_h)	-	\mathcal{A}_0(\zeta_C)	\Big) 
			\\&
			+	\frac{1	}{(\dimd-1)(\dimd-2)}	\Big( \mathcal{A}_0(\zeta_h)	-	\mathcal{A}_0(\zeta_R)	\Big) 
			\bigg]
			\,\text{.}
		\end{aligned}
		\label{eq:amplitudefunctional}
	\end{equation}
	Here we have defined the partial Green's functionals
	\begin{align}
		\mathcal{A}_0(\zeta)	&=	\int \dd[\dimd]{x}\sqrt{-g}	\,	V[\chi_1,\chi_2]	\,	\mathcal{G}_0(\square;\zeta)	\,	V[\phi_1,\phi_2]
		\,\text{,}\\
		\mathcal{A}_2(\zeta)	&=	\int \dd[\dimd]{x}\sqrt{-g}	\,	T_{\alpha\beta}[\chi_1,\chi_2]	\,	\mathcal{G}_2(\square;\zeta)	\,	T^{\alpha\beta}[\phi_1,\phi_2]
		\,\text{,}
	\end{align}
	the dimensionless mass parameters
	\begin{equation}\begin{aligned}
			\zeta_R &=	\frac{1}{\aRR H^2}	+	\frac{1}{2}\dimd(\dimd-4) +	\frac{\aE}{\aRR}
			\,\text{,}&\qquad&
			\zeta_h = -2(\dimd-1)
			\,\text{,}\\
			\zeta_C &= 	\frac{1}{\aCC H^2}	-	\dimd	+	\frac{1}{2}\dimd(\dimd-2)	\frac{\aRR}{\aCC}	+	\frac{\aE}{\aCC}	
			\,\text{.}
		\end{aligned}\label{eq:masseszeta}\end{equation}
	and the coefficient
	\begin{equation}
		c(\aRR,\aE) = \frac{1}{1+\aE H^2}\frac{\zeta_R-\frac{1}{2}\dimd(\dimd-4)}{\zeta_R+\dimd}
		\,\text{.}
	\end{equation}

	\subsection{Discussion}\label{sec:amplitudefunctionaldiscussion}
	From the Green's functional \eqref{eq:amplitudefunctional}, the following observations can be made straightaway.
	
	First, we note that the Green's function does not depend on the gauge parameters $\alphagf$ and $\betagf$. This shows that the on-shell Green's function is gauge fixing independent, conform the expectation that the Green's function (and hence its related scattering amplitude) is related to an observable.
	
	Second, we note that the scalar vertex $V[\phi_1,\phi_2]$ vanishes for a conformally coupled scalar field $m_\phi =0$, $\aR{\phi}= \frac{\dimd-2}{4(\dimd-1)}$. This greatly simplifies the Green's function. For this reason, the conformally coupled scalar field has been studied widely \cite{Tsamis:1992xa,Park:2015kua,Frob:2016fcr, Frob:2017smg,Glavan:2020ccz}.
	
	Third, we observe that the Green's functional \eqref{eq:amplitudefunctional} depends on three mass parameters $\zeta_h$, $\zeta_R$ and $\zeta_C$. In the limit $H\to 0$, these correspond to masses of zero, $\aRR^{-1}$ and $\aCC^{-1}$, respectively. Hence, these mass parameters correspond to the massless graviton, and the spin-0 and spin-2 mass poles. At first inspection, the negative value of $\zeta_h$ may seem to yield a tachyonic particle. However, it is important to keep in mind that due to the nonzero curvature the correspondence between the sign of the mass and the causal behavior is not obvious \cite{Garidi:2003ys, Pejhan:2018ofn}.
	
	Fourth, it is now relatively simple to obtain the Green's functions of GR, $R^2$-gravity and $C^2$-gravity from this expression, by taking the following limits:
	\begin{align}
		\text{GR:}	&\qquad	\aCC\to0, \aRR\to 0 \,\text{;}\label{eq:GRlimit}
		\\
		\text{$R^2$-gravity:}	&\qquad	\aCC\to 0	\,\text{;}\label{eq:R2limit}
		\\
		\text{$C^2$-gravity:}	&\qquad	\aRR\to 0	\,\text{.}\label{eq:C2limit}
	\end{align}
	Furthermore, we obtain the Green's function in Minkowski spacetime by taking the limit
	\begin{equation}
		\text{Minkowski spacetime:} \qquad H \to 0
		\,\text{.}
	\end{equation}
	Let us first consider the latter limit. We note that the mass parameters $\zeta_C$ and $\zeta_R$ go to infinity; we then find the following limits for the coefficients:
	\begin{equation}
		\lim_{H\to0} c(\aRR,\aE) 
		=	\lim_{H\to0} \frac{\zeta_C+\dimd}{\zeta_C-\zeta_h} 
		=	\lim_{H\to0} \frac{\zeta_C}{\zeta_C-\zeta_h}
		= 1
		\,\text{.}
	\end{equation}
	The propagators reduce to
	\begin{equation}\label{eq:propagatormasslesslimit}
		\begin{aligned}
			\lim_{H\to0}	\mathcal{G}_0(\square;\zeta_h) &= \lim_{H\to0}	\mathcal{G}_2(\square;\zeta_h) = \square^{-1}
			\,\text{,}\\
			\lim_{H\to0}	\mathcal{G}_0(\square;\zeta_C) &= \lim_{H\to0}	\mathcal{G}_2(\square;\zeta_C) = \Big( \square + \aCC^{-1} \Big)^{-1}
			\,\text{,}\\
			\lim_{H\to0}	\mathcal{G}_0(\square;\zeta_R) &= \Big( \square + \aRR^{-1} \Big)^{-1}
			\,\text{.}
		\end{aligned}
	\end{equation}
	This is consistent with the amplitude found in \cite{Stelle:1976gc,Donoghue:1994dn}. Inspecting the propagators \eqref{eq:propagatormasslesslimit}, we see that the Minkowski spacetime amplitude corresponds to a massless spin-0 and spin-2 particle, corresponding to the graviton, and in addition a massive spin-0 particle of mass $\aRR^{-1}$ and a massive spin-2 particle of mass $\aCC^{-1}$ \cite{Stelle:1976gc, Stelle:1977ry}.
	
	In a similar fashion, we consider $R^2$- and $C^2$-gravity. Taking the limit $\aRR \to 0$ and $\aCC \to 0$, we find that $\zeta_R$ and $\zeta_C$ are linearly divergent, respectively. Hence, provided that the coefficients of the propagators remain finite, we conclude that the propagators $\mathcal{G}_0(\square;\zeta_C)$ and $\mathcal{G}_2(\square;\zeta_C)$ are suppressed in $R^2$-gravity, corresponding to a decoupling of the spin-2 Stelle particle, while in $C^2$-gravity the propagator $\mathcal{G}_0(\square;\zeta_R)$ is suppressed, corresponding of a decoupling of the spin-0 massive particle from the theory.
	It remains to check that the coefficients of the propagator stay finite; we compute
	\begin{equation}\begin{aligned}
			\lim_{\aRR\to0} c(\aRR,\aE)  = \frac{1}{1+\aE H^2}
			\,\text{,}&\qquad&
			\lim_{\aCC\to0} \frac{\zeta_C+\dimd}{\zeta_C-\zeta_h} 
			=	\lim_{\aCC\to 0} \frac{\zeta_C}{\zeta_C-\zeta_h}
			= 1
			\,\text{.}
	\end{aligned}\end{equation}
	This justifies the assertion that the massive particles decouple in $R^2$- and $C^2$-gravity.
	
	\section{The Green's function in the adiabatic limit}\label{sec:adiabatic-amplitude}
	We will now inspect the Green's functional in more detail. Since the Green's functional is given by abstract differential operators on \deS{} spacetime, it is not easy to draw conclusions about its physical properties. Therefore, we  resort to expansion methods to turn \eqref{eq:amplitudefunctional} into a concrete numerical expression. We employ an expansion around the scalar masses $\mu = m/H = \infty$, following \cite{Ferrero:2021lhd}. This is a specific realization of the so-called adiabatic expansion \cite{Agullo:2014ica,Moreno-Pulido:2022phq, Junker:2001gx, Lueders:1990np, Olbermann:2007gn, Fulling1979RemarksOP, Fulling:1974zr, Fulling:1989nb, Parker:1968mv, Parker:1969au, Parker:1971pt, Parker:1974qw, Parker:2009uva, Birrell_adiabatic, Bunch_adiabatic, Bunch:1978aq, Haro:2008zz, Winitzki:2005rw, Zeldovich1,Zeldovich2}. This can be understood from the observation that an expansion around $\mu =\infty $ implies that the evolution of \deS{} spacetime is much slower than the Compton frequency of the scalar fields.
	
	In \autoref{sec:adiabatic-amplitude-computation}, we describe the algorithm used to compute the adiabatic limit of the Green's functional. This computation was pioneered in the context of GR in \cite{Ferrero:2021lhd}; here we generalize the calculation to \QuadG.	
	The properties of the resulting Green's function are studied in \autoref{sec:adiabatic-amplitude-analysis}. In \autoref{app:adiabatic}, we collect several details regarding the adiabatic limit obtained in \cite{Ferrero:2021lhd}. \autoref{app:details} is devoted to the  numerical techniques to obtain the Green's function in the adiabatic limit.
	
	\subsection{Computation of the Green's function}\label{sec:adiabatic-amplitude-computation}
	Let us start by evaluating the wave function in a Bunch-Davies vacuum with outgoing momenta.
	Explicitly, these mode functions are given by	\begin{align}
		\phi_{\vec p,\mu}(\eta,\vec x) = \eta^{\frac{\dimd-1}{2}}	H^{(1)}_{\imath\sqrt{\mu^2-\left(\frac{\dimd-1}{2}\right)^2}}(-|\vec{p}|\eta) \ee^{\imath \vec p \cdot \vec x}
		\,\text{.}
\end{align}
	
Subsequently, by invoking the Bunch-Davies vacuum, we can formalize the adiabatic expansion. The adiabatic limit is implemented by performing an expansion in the dimensionless parameters
	\begin{equation}
		\begin{aligned}
			\mu_\phi = m_\phi/H
			\,\text{,}&\qquad&
			\mu_\chi = m_\chi/H
			\,\text{.}
		\end{aligned}
	\end{equation}
	around infinity. 
	Using the results in \autoref{app:adiabatic}, we are able to expand the solutions to the wave equation in powers of $\mu_\phi$ and $\mu_\chi$. We use this evaluate the Green's functional.
	We observe that in order to expand \eqref{eq:amplitudefunctional}, it suffices to compute the adiabatic limit of $\mathcal{A}_0$ and $\mathcal{A}_2$. 
	Following the argument in \autoref{app:details}, we find that the leading terms are proportional to $\mu_\phi^2 \mu_\chi^2$. Thus, we define:
	\begin{equation}\begin{aligned}
			\mathcal{A}_0 &= \int \dd[\dimd]{x}\sqrt{-g}\, A_0(\proper{q};\zeta) \, \chi_1 \chi_2\phi_1\phi_2	+	\order{\mu_\chi,\mu_\phi}
			\,\text{,}\\
			\mathcal{A}_2 &=   \int\dd[\dimd]{x}\sqrt{-g}\, A_2(\proper{q};\zeta)	\,\chi_1 \chi_2\phi_1\phi_2	+	\order{\mu_\chi,\mu_\phi}
			\,\text{.}
	\end{aligned}\end{equation}
	Here the Green's  functionals $\mathcal{A}_i$ are determined by the integration kernels $A_i(\proper{q};\zeta)$. These are functions of the proper momentum transfer $\vec{\proper{q}} = -\eta \vec{q} = -\eta \left(\vec{p}_1 - \vec{p}_2\right)$, where $\vec p_1$ is the momentum of $\phi_1$ and $\vec p_2$ is the momentum of $\phi_2$.

	We now discuss the general strategy to compute $A_i$. In abuse of terminology, we refer to the $A_i$ as amplitudes. While all expressions in \autoref{sec:amplitudefunctional} are valid in any coordinate basis, we will now work explicitly in conformally flat coordinates. This allows to go to momentum space for spatial derivatives.
	
	In these coordinates, we can assert from covariance that the propagator $\mathcal{G}_0(\square,\zeta)$ acting on the product $\phi_1\phi_2$ can be parameterized by
	\begin{equation}
		\mathcal{G}_0(\square;\zeta)	\phi_1 \phi_2 = \left[	G_0(\proper{q};\zeta)	+	\order{\mu_\phi^{-1}}	\right]	\phi_1\phi_2
		\,\text{,}
	\end{equation}	
	for some function $G_0$. The latter is found as follows.
	Acting with the inverse propagator $\mathcal{G}_0^{-1}(\square;\zeta)=\left(\square+\zeta H^2\right)$ should give the identity operator. Distributing the derivatives gives the second-order inhomogeneous differential equation\footnote{An equation of this type was also derived and solved with similar techniques in \cite{Goodhew:2022ayb}.}
	\begin{equation}\label{eq:G0equation}
		\eta^2	\partial_{\eta}^2G_0	+	\dimd \eta \partial_\eta  G_0+	(q^2\eta^2	+\zeta)G_0	=	\frac{1}{H^2}
		\,\text{.}
	\end{equation}
Now we can perform a change of variables and express the propagator as a function of $\proper{q}$.
	Following the argument in \autoref{app:details}, it is natural to  require that $G_0$ is analytic in $\proper{q}$ for any dimension $\dimd$. Since the homogeneous solution to \eqref{eq:G0equation} is in general not analytic in $\proper{q}$, $G_0$ is simply given by the inhomogeneous solution:
	\begin{equation}
		G_0(\proper{q};\zeta)	=	\frac{1}{\zeta H^2}	\HypergeometricPFQtilde{1}{2}{1}{\frac{\dimd+3}{4},\frac{\nu(\zeta)}{2}}{-\frac{\proper{q}^2}{4}}
		\,\text{.}
	\end{equation}
	Here we defined $\HypergeometricPFQtilde{1}{2}{a}{b,c}{z} = \HypergeometricPFQ{1}{2}{a}{b-c,b+c}{z}$ in terms of a generalized hypergeometric function. In addition, we defined the parameter
	\begin{equation}
		\nu(\zeta) = \sqrt{\left( \frac{\dimd-1}{2}	\right)^2-\zeta}
		\,\text{.}
	\end{equation}
	Here a remark is in order. In Minkowski space different propagators are chosen,  with different complex structures $i \varepsilon$ corresponding to different boundary conditions (Feynman, retarded, advanced). For our purposes, constructing the Green's function by acting with the propagator on heavy-mass particles states, the complex structure is not relevant. In fact, the same is valid also for the flat spacetime case when the Born approximation is invoked.

	We thus find
	\begin{equation}
		A_0(\proper{q};\zeta) = \frac{\mu_\phi^2\mu_\chi^2 H^2}{\zeta}	\HypergeometricPFQtilde{1}{2}{1}{\frac{\dimd+3}{4},\frac{\nu(\zeta)}{2}}{-\frac{\proper{q}^2}{4}}
		\,\text{.}
		\label{eq:A0}
	\end{equation}
	This completes our calculation of $A_0$.
	
	For $A_2$, we  take into account the tensor structure of $T_{\alpha\beta}$. Due to the two-derivative structure of $T_{\alpha\beta}$, the leading order will be of order $\mu_\phi^2$ as shown in \autoref{app:details}.  We then make the following ansatz for the propagator acting on $T_{\alpha\beta}$:
	\begin{equation}\begin{aligned}
			\mathcal{G}_2(\square;\zeta)	T_{00}[\phi_1,\phi_2]	&=	\big[	G_{00}(\proper{q})\mu_\phi^2	+	\order{\mu_\phi}	\big]	\phi_1\phi_2
			\,\text{,}\\
			\mathcal{G}_2(\square;\zeta)	T_{0i}[\phi_1,\phi_2]	&=	\big[	\big(	G_{1}(\proper{q})p_{1,i}	+	G_{2}(\proper{q})p_{2,i}	\big)	\mu_\phi^2	+	\order{\mu_\phi}	\big]	\phi_1\phi_2
			\,\text{,}\\
			\mathcal{G}_2(\square;\zeta)	T_{ij}[\phi_1,\phi_2]	&=	\begin{aligned}[t]&
				\big[	\big(	G_{(12)}(\proper{q})	\left(p_{1,i}p_{2,j}+p_{1,j}p_{2,i}\right)	+	G_\delta(\proper{q})\delta_{ij}
				\\	
				&\quad+	G_{11}(\proper{q})p_{1,i}p_{1,j}	+	G_{22}(\proper{q})p_{2,i}p_{2,j}	\big)	\mu_\phi^2	+	\order{\mu_\phi}	\big]	\phi_1\phi_2
				\,\text{.}		\end{aligned}
	\end{aligned}\end{equation}
	Acting on these expressions with $\mathcal{G}_2^{-1}(\square;\zeta)= \left(\square+(\zeta+2\dimd) H^2\right)$ gives a set of coupled differential equations. 
	Using the expansion above, we find
	\begin{multline}
		A_2(\proper{q};\zeta)
		=
		\frac{H^4\eta^2}{\dimd}	\mu_\phi^2 \mu_\chi^2	\Big[	
		-	p_1^2 G_{11}(\proper{q})	-	p_2^2 G_{22}(\proper{q})	-2	p_1\cdot p_2 \, G_{(12)}(\proper{q})	
		\\
		(\dimd-1)	\left(	G_{00}(\proper{q})	-	G_{\delta}(\proper{q})	\right)	
		\Big]
		\,\text{.}
	\end{multline}
	By making linear combinations $g_i(\proper{q}) = \sum_{j}a_{ij}G_j(\proper{q})$, the system can be partially decoupled.
	Since the exact linear combination is rather complicated, we will refrain from reproducing it here. We refer to \cite{Ferrero:2021lhd} and the accompanying notebook of that reference for  details on how to compute this. At this point, it suffices to note that $A_2$ is given by
	\begin{equation}
		A_2(\proper{q};\zeta)
		=
		\frac{\mu_\phi^2 \mu_\chi^2 H^2	\proper{q}^2}{2\dimd}	\Big(	(\dimd-2)g_1(\proper{q}) +	\dimd g_2(\proper{q})	\Big)	
		\,\text{.}
	\end{equation}
	Here the functions $g_1$ and $g_2$ satisfy the differential equations
	\begin{equation}\label{eq:g123}
		\begin{aligned}
			\proper{q}^2	g_1''	+	(\dimd+4)	\proper{q}	g_1'
			&+	(\proper{q}^2+\zeta+3\dimd+2)	g_1
			=
			\frac{1}{\proper{q}^2}
			+	\dimd g_2
			\,\text{;}\\
			\proper{q}^2	g_2''	+	(\dimd+4)	\proper{q}	g_2'
			&+	(\proper{q}^2+\zeta+3\dimd) g_2
			=
			\frac{1}{\proper{q}^2}
			+	(\dimd-2)g_1
			+	4\imath \proper{q} g_3
			\,\text{;}\\
			\proper{q}^2	g_3''	+	(\dimd+4)	\proper{q}	g_3'
			&+	(\proper{q}^2+\zeta+3\dimd) g_3
			=
			4	\imath \proper{q} g_2
			\,\text{.}
		\end{aligned}
	\end{equation}
	We solve this system of equations by inserting a power series ansatz \cite{Frobenius1873}:
	\begin{equation}
		g_i(\proper{q)} = \proper{q}^{\nu_i} \sum_{j\geq0}	b_{ij} \proper{q}^j
		\,\text{.}
	\end{equation}
	Requiring that $A_2$ is analytic in $\eta$ then implies that $\nu_i \in \mathbb{Z}$, and that the $g_i$ are fixed by the inhomogeneous solution to \eqref{eq:g123}. Inserting this ansatz into the differential equations \eqref{eq:g123} allows to resolve $\nu_i$ and the coefficients $b_{ij}$, after which we can resum the power series. The solution can be written as
	\begin{equation}
		A_{2}(\proper{q};\zeta) = A_{2}^{\text{reg}}(\proper{q};\zeta)	+	A_{2}^{\text{sing}}(\proper{q};\zeta)
		\,\text{.}\label{eq:A2tot}
	\end{equation}
	Here the term $A_{2}^{\text{reg}}$ is regular for $\zeta=\zeta_h$:
	\begin{equation}\label{eq:A2regular}
		\begin{aligned}		A_{2}^{\text{reg}}(\proper{q};\zeta)
			=		&
			\frac{\mu_\phi^2 \mu_\chi^2 H^2}{\zeta+\dimd}	\bigg[
			\frac{1}{\dimd-1}	\HypergeometricPFQtilde{1}{2}{1}{\frac{\dimd-1}{4},\frac{\nu(\zeta)}{2}}{-\frac{\proper{q}^2}{4}}
			\\&
			+	\left[\frac{\dimd-3}{\zeta}-\frac{1}{\dimd}\right]	\HypergeometricPFQtilde{1}{2}{1}{\frac{\dimd+3}{4},\frac{\nu(\zeta)}{2}}{-\frac{\proper{q}^2}{4}}
			\\&
			+	\frac{2}{\zeta}	\HypergeometricPFQtilde{1}{2}{2}{\frac{\dimd+3}{4},\frac{\nu(\zeta)}{2}}{-\frac{\proper{q}^2}{4}}
			\bigg]
			\,\text{.}	
		\end{aligned}
	\end{equation}
	Special care has to be taken into the case $\zeta = \zeta_h$. This is summarized in the function  $A_2^{\text{sing}}(\proper{q};\zeta)$, which is given by
	\begin{equation}\label{eq:A2singular}
		A_{2}^{\text{sing}}(\proper{q};\zeta \neq \zeta_h)=\frac{\mu_\phi^2 \mu_\chi^2 H^2}{\zeta+\dimd}	\frac{\dimd-2}{\dimd-1}		\HypergeometricPFQtilde{1}{2}{1}{\frac{\dimd-1}{4},\frac{\nu(\zeta-\zeta_h)}{2}}{-\frac{\proper{q}^2}{4}}
	\end{equation}
	for $\zeta \neq \zeta_h$. For the special value $\zeta = \zeta_h$ we find
	\begin{equation}\label{eq:A2singularh}
		\begin{aligned}
			A_{2}^{\text{sing}}(\proper{q};\zeta = \zeta_h)=	-&	\frac{\mu_\phi^2 \mu_\chi^2 H^2}{2}	\bigg[ 
			\HypergeometricZeroFOne{\frac{\dimd+1}{2}}{-\frac{\proper{q}^2}{4}}
			\\&
			-\frac{(\dimd-3)(61+252\dimd-14\dimd^2-12\dimd^3+\dimd^4)}{12(\dimd-1)^2(\dimd+1)(\dimd+5)}	\HypergeometricZeroFOne{\frac{\dimd-3}{2}}{-\frac{\proper{q}^2}{4}}
			\\&
			+	\frac{(\dimd-3)(-11-26\dimd+\dimd^2)(-11+2\dimd+\dimd^2)}{12(\dimd-1)^2(\dimd+1)(\dimd+5)}	\HypergeometricZeroFOne{\frac{\dimd-1}{2}}{-\frac{\proper{q}^2}{4}}
			\\&
			+	\frac{2}{	\dimd-1}	\bigg(
			\HypergeometricPFQDerivative{(0;0,1;0)}{1}{2}{1}{1,\frac{\dimd-1}{2}}{-\frac{\proper{q}^2}{4}}
			\\&\hspace{6em}
			-	\HypergeometricPFQDerivative{(0;0,1;0)}{1}{2}{2}{1,\frac{\dimd-1}{2}}{-\frac{\proper{q}^2}{4}}
			\bigg)
			\\&
			-	\frac{2}{	\dimd-1}	\bigg(
			\HypergeometricPFQDerivative{(0;1,0;0)}{1}{2}{1}{1,\frac{\dimd-1}{2}}{-\frac{\proper{q}^2}{4}}
			\\&\hspace{6em}
			-	\HypergeometricPFQDerivative{(0;1,0;0)}{1}{2}{2}{1,\frac{\dimd-1}{2}}{-\frac{\proper{q}^2}{4}}
			\bigg)
			\bigg] \,\text{.}
		\end{aligned}
	\end{equation}
	This completes the computation of $A_0$ and $A_2$. The total Green's function can now be written as
	\begin{equation}\label{eq:amplitudeQuadG}
		\begin{aligned} A(\proper{q}) = 16\pi \GN \, c(\aRR,\aE)	\bigg[ &
			-	\frac{\zeta_C+\dimd}{\zeta_C-\zeta_h}	\Big(	A_2(\zeta_h)	-	A_2(\zeta_C)	\Big)
			\\&
			+	\frac{1}{\dimd(\dimd-1)}	\frac{\zeta_C}{\zeta_C-\zeta_h}	\Big( A_0(\zeta_h)	-	A_0(\zeta_C)	\Big) 
			\\&
			+	\frac{1	}{(\dimd-1)(\dimd-2)}	\Big( A_0(\zeta_h)	-	A_0(\zeta_R)	\Big) 
			\bigg]
			\,\text{.}
		\end{aligned}
	\end{equation}
	In \eqref{eq:amplitudeQuadG}, $A_2(\zeta_h)$ is given by \eqref{eq:A2singularh},  $A_2(\zeta_C)$ by \eqref{eq:A2singular}, and $A_0(\zeta)$ by \eqref{eq:A0}.
	
	\subsection{Analysis of the Green's function}\label{sec:adiabatic-amplitude-analysis}
	In this section, we study the properties of the Green's function computed above. For simplicity, we will limit ourselves to the case $\dimd =4$ only. We will first list several features of the amplitudes in \autoref{sec:amplitudeproperties}. In \autoref{sec:amplitudeinterpretation}, we discuss their interpretation.

	\subsubsection{Properties of the \texorpdfstring{$A_i$}{Ai}}\label{sec:amplitudeproperties}
\begin{figure}
		\centering
		\begin{subfigure}{.5\textwidth}
			\includegraphics[width=\textwidth]{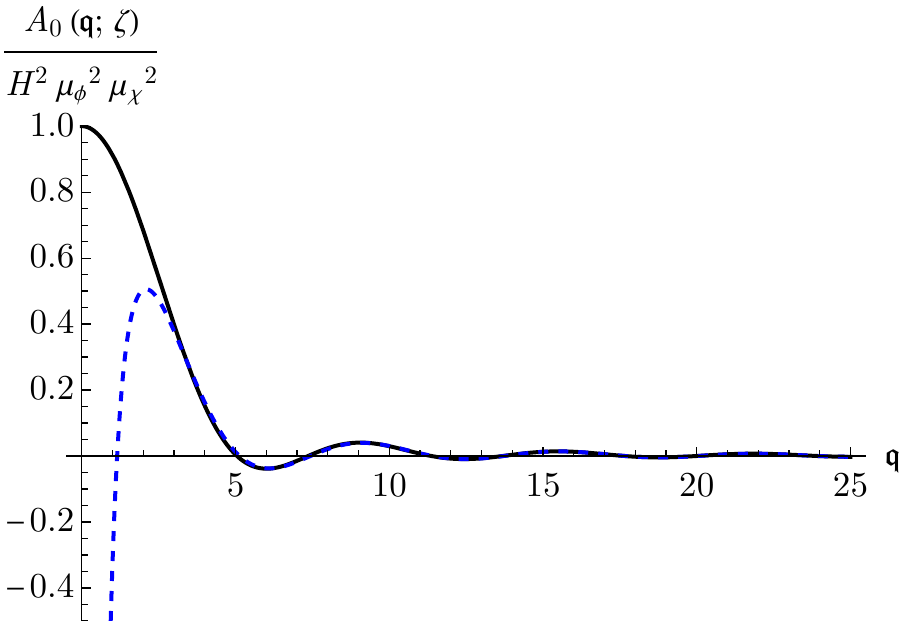}
			\caption{}
		\end{subfigure}%
		\begin{subfigure}{.5\textwidth}
			\includegraphics[width=\textwidth]{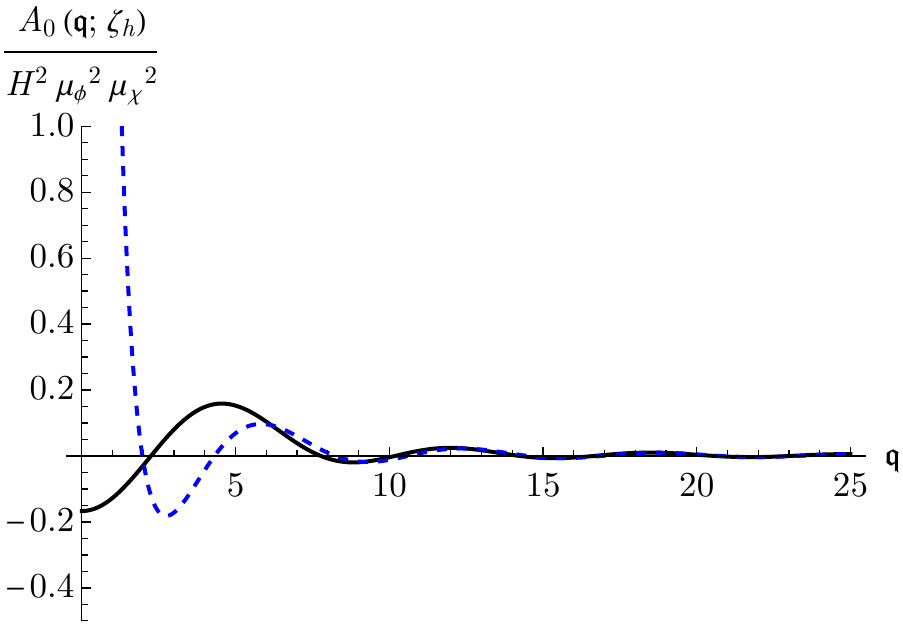}
			\caption{}
		\end{subfigure}
		\caption{The amplitude $A_0(\proper{q};\zeta)$ in $\dimd=4$ for several values of $\zeta$. Left panel: $A_0(\zeta=1)$. Right panel: $A_0(\zeta=\zeta_h)$. For $\proper{q}=0$, the amplitude attains the finite value given in \eqref{eq:A0q0}. For large $\proper{q}$, the amplitude behaves as $\sim\proper{q}^{-2}\left(1- \alpha \cos(\proper{q})\right)$, as computed in \eqref{eq:A0qinfty}. This approximation is depicted by the dashed blue lines.
		}
		\label{fig:A0}
	\end{figure}

\begin{figure}
		\centering
		\begin{subfigure}{.5\textwidth}
			\includegraphics[width=\textwidth]{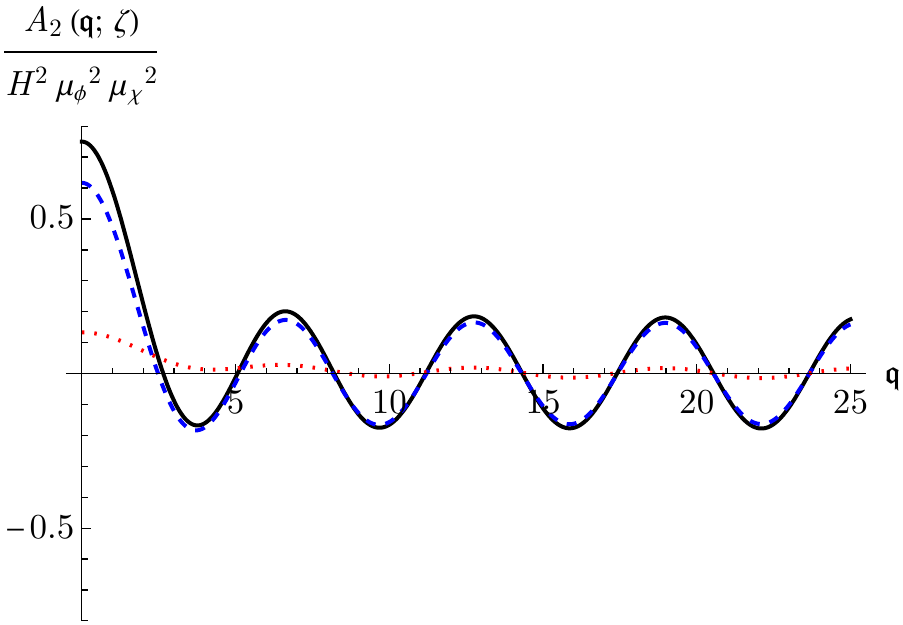}
			\caption{}
		\end{subfigure}%
		\begin{subfigure}{.5\textwidth}
			\includegraphics[width=\textwidth]{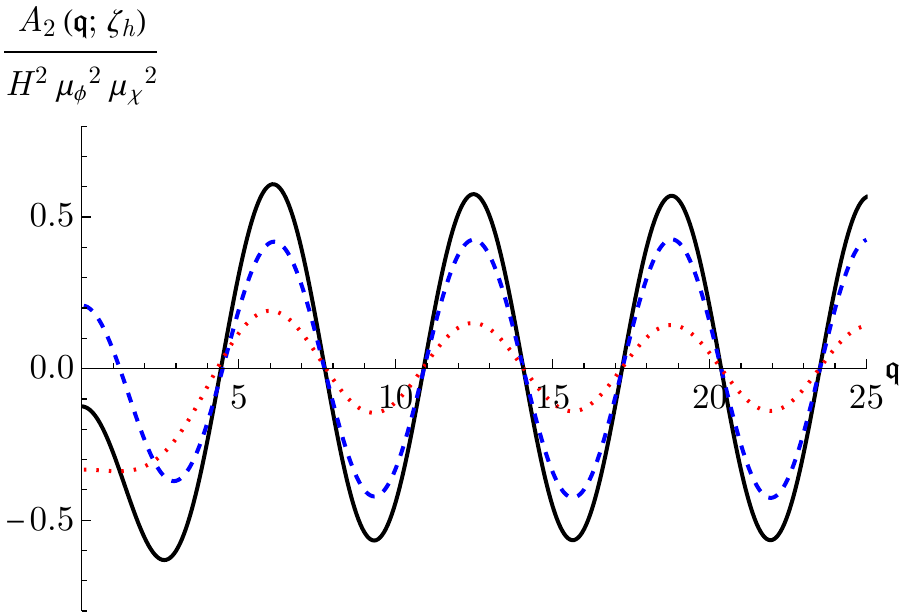}
			\caption{}
		\end{subfigure}
		\caption{The amplitude $A_2(\proper{q};\zeta)$ in $\dimd=4$ for several values of $\zeta$. Left panel: $A_2(\zeta=1)$. Right panel: $A_2(\zeta=\zeta_h)$. Solid line: the total amplitude $A_2$. Dashed blue line: $A_2^{\text{reg}}$. Dotted red line: $A_2^{\text{sing}}$. For $\proper{q}=0$, the amplitude attains the finite values given by \eqref{eq:A2regq0} and \eqref{eq:A2singq0}. For large $\proper{q}$, the amplitude behaves as $\sim \cos(\proper{q})$, as computed in \eqref{eq:A2regqinfty} and \eqref{eq:A2singqinfty}.
		}
		\label{fig:A2}
	\end{figure}

	In order to get a feeling for the behavior of the $A_i$, we plot the amplitude $A_0$ in \autoref{fig:A0} for several values of $\zeta$. In \autoref{fig:A2}, the amplitude $A_2$ is shown.
	
	We distinguish two features of the amplitude. First, we note that since the hypergeometric function ${}_1\hskip-1pt F_2$ is bounded, the amplitude is bounded too. In particular, it attains a finite value for $\proper{q}\to 0$. For $A_0$, we find
	\begin{equation}\label{eq:A0q0}
		\frac{A_0(\proper{q}=0;\zeta)}{H^2\mu_\phi^2\mu_\chi^2} = \frac{1}{\zeta}
		\,\text{.}
	\end{equation}
	For $A_2$, we first consider the regular part $A_2^{\text{reg}}$. The limit $\proper{q}\to 0$ reads
	\begin{equation}\label{eq:A2regq0}
		\frac{A_2^{\text{reg}}(\proper{q}=0;\zeta)}{H^2\mu_\phi^2\mu_\chi^2} = 
		\frac{2}{3}\frac{1}{\zeta+4}
		+	\frac{3}{4}\frac{1}{\zeta}
		\,\text{,}
	\end{equation}
	while for the singular part $A_2^{\text{sing}}$, we compute
	\begin{equation}\label{eq:A2singq0}
		\frac{A_2^{\text{sing}}(\proper{q}=0;\zeta)}{H^2\mu_\phi^2\mu_\chi^2} =
		\frac{2}{3}	\frac{1}{\zeta+4}
		\,\text{,}
	\end{equation}
	holding both for $\zeta \neq \zeta_h$ and for $\zeta = \zeta_h$.
	
	The second feature that we study is the behavior at large $\proper{q}$. Expanding around $\proper{q}=\infty$, we obtain for $A_0$:
	\begin{equation}\label{eq:A0qinfty}
		\frac{A_0(\proper{q};\zeta)}{H^2\mu_\phi^2\mu_\chi^2} \sim 
		\frac{1}{\proper{q}^2}
		+	\frac{4}{\sqrt{\pi}\zeta}	\Gamma\left(\tfrac{7}{4}-\tfrac{\nu(\zeta)}{2}\right)\Gamma\left(\tfrac{7}{4}+\tfrac{\nu(\zeta)}{2}\right)	\frac{1}{\proper{q}^2}	\cos\left(\proper{q}\right)
		\,\text{.}
	\end{equation}
	Expanding $A_2^{\text{reg}}$ around large momentum transfer, we find
	\begin{equation}\label{eq:A2regqinfty}
		\frac{A_2^{\text{reg}}(\proper{q},\zeta)}{H^2\mu_\phi^2\mu_\chi^2}
		\sim
		-	\frac{1}{3\sqrt{\pi}}	\frac{\Gamma\left(\tfrac{3}{4}-\tfrac{\nu(\zeta)}{2}\right)\Gamma\left(\tfrac{3}{4}+\tfrac{\nu(\zeta)}{2}\right)}{\zeta+4}	\cos\left(\proper{q}\right)
		\,\text{.}
	\end{equation}
	The large-$\proper{q}$ expansion of $A_2^{\text{sing}}$ reads
	\begin{equation}\label{eq:A2singqinfty}
		\frac{A_2^{\text{sing}}(\proper{q};\zeta)}{H^2\mu_\phi^2\mu_\chi^2} \sim \begin{cases}
			\begin{aligned}
				\frac{2}{3\sqrt{\pi}}\frac{\Gamma\left(\tfrac{3}{4}-\tfrac{\nu(\zeta-\zeta_h)}{2}\right)\Gamma\left(\tfrac{3}{4}+\tfrac{\nu(\zeta-\zeta_h)}{2}\right)}{\zeta+4}  \cos\left(\proper{q}\right)
			\end{aligned}	\,\text{,}& \zeta\neq\zeta_h
			\\ \\
			\begin{aligned}
				\frac{397-360 \log(2)}{1080}	\cos\left(\proper{q}\right)
			\end{aligned}\,\text{,}& \zeta=\zeta_h
		\end{cases}\,\text{.}
	\end{equation}
	We therefore find the following behavior: for large $\proper{q}$, the amplitudes are oscillating. Restoring dimensionful quantities, we find that the oscillating terms are suppressed by $H^2$. Carefully taking the limit $H\to 0$, which we will not further discuss here, allows to recover the flat-spacetime $1/\vec{q}^2$ behavior.
	
	\subsubsection{Interpretation of the amplitudes}\label{sec:amplitudeinterpretation}
	We will now briefly discuss the interpretation of the limits $\proper{q}\to 0$ and $\proper{q}\to \infty$.
	
	Let us first consider the small-momentum limit, which we find to be finite.	This behavior is in line with the expectation that $z=\zeta H^2$ acts as a mass, similar to the limit $\vec{q} \to 0$ for a flat-spacetime massive propagator $\propto \frac{1}{\vec{q}^2 + z}$. A special case arises when $\zeta=\zeta_h$. Then \eqref{eq:A0q0}, \eqref{eq:A2regq0} and \eqref{eq:A2singq0} are manifestly finite. In flat spacetime, on the other hand, we have a massless propagator, $\frac{1}{\vec{q}^2}$, which is divergent in the limit $\vec{q} \to 0$. We therefore conclude that the background curvature acts as an infrared regulator.
	
	The oscillations in the large-momentum limit were also observed in \cite{Ferrero:2021lhd} in the context of GR. As discussed there, the oscillations are typical for propagators in \deS{} spacetime \cite{Arkani-Hamed:2015bza}. Furthermore, the discrete values where the amplitude vanishes has an interesting interpretation in terms of a probability density. A node at momentum $\proper{q}_0$ then implies that the exchange of a graviton with this momentum is forbidden. These momenta are equidistantly separated with distance $\Delta \proper{q}=2\pi$. This discrete behavior is reminiscent of discrete transition probabilities associated to particles in a box. In this context, the Hubble volume then acts as the boundary within which gravitons can propagate.
	
	\section{Scattering potential in the adiabatic expansion}\label{sec:potential}
	The amplitude $A(\proper{q})$ represents the transition probability of the scattering process $\phi \chi \to \phi \chi$ with momentum transfer $\proper{q}$. Converting to position space, we find the transition amplitude of the same scattering process, where the external states are now localized at well-determined spacetime positions. As a generalization of the Born approximation in flat spacetime, we interpret this object as the scattering potential.
	
	\subsection{Computation of the scattering potential}
	We obtain the transition amplitude in position space by taking the Fourier transform of \eqref{eq:amplitudeQuadG}. Preparing particle $\phi_1$ at position $\vec{x}_1$ and particle $\chi_1$ at position $\vec{x}_2$, the transition probability arising from the amplitude $A_i$ is given by
	\begin{equation}
		V_i(\vec x_1,\vec x_2) 
		= 
		\frac{1}{2 \mu_\phi \mu_\chi}	\int \frac{\dd[\dimd-1]{\vec{\proper{k}}_1}}{(2\pi)^{\dimd-1}}	\frac{\dd[\dimd-1]{ \vec{\proper{p}}_1}}{(2\pi)^{\dimd-1}}	\ee^{\imath \vec p_1 \cdot \vec x_1} \ee^{\imath \vec k_1 \cdot \vec x_2}	A_i(\proper{q;\zeta})
		\,\text{.}
	\end{equation}
	Here the dimensionless proper momenta $\vec{\proper{p}}_1 = - \eta \vec{p}_1$ and $\vec{\proper{k}}_1$ such that $\vec{q} = \vec{p}_1-\vec{k}_1$ are integrated over to obtain an invariant expression. In addition, the prefactor of the integral is chosen such that the resulting potential is correctly normalized.
	
	Performing the integral over $\vec{\proper{k}}_1$ and the spherical part of $\vec{\proper{p}}_1$, we find the dimensionless potential:
	\begin{align}
		V_i(\vec x_1,\vec x_2)	&=	\delta^{\dimd-1}(H\vec x_2)	V_i(\proper{r})
		\,\text{,}\\
		V_i(\proper{r})	&=	\frac{1}{2 \mu_\phi \mu_\chi}	\frac{2^{2-\dimd}	\pi^{\frac{1-\dimd}{2}}}{\Gamma\left(\frac{\dimd-1}{2}\right)}	\int_0^\infty	\dd{\proper{q}}		\proper{q}^{\dimd-2}	\HypergeometricZeroFOne{\frac{\dimd-1}{2}}{-\frac{\proper{q}^2\proper{r}^2}{4}}	A_i\left(\proper{q}\right) \label{eq:potentialfouriertransform}
		\,\text{.}
	\end{align}
	Here, $\proper{r} = \lVert \eta(\vec{x}_1-\vec{x}_2)\rVert$ denotes the proper distance between the particles $\phi_1$ and $\chi_1$.\footnote{The distance is computed using the $\delta_{ij}$, being the spatial part of the Poincaré metric with the conformal factor incorporated.}
	Central in this computation is the integral
	\begin{equation}\label{eq:finitedifference}
		\int_0^\infty	\dd{x} x^{s-1} \HypergeometricZeroFOne{\alpha}{-ax} \HypergeometricPFQ{1}{2}{n}{\beta_1,\beta_2}{-bx}
		\,\text{.}
	\end{equation}
	This has the structure of a Mellin transform, and was studied in \cite{MILLER2001}. In order to compute the Fourier transform of the functions $\HypergeometricPFQDerivative{(0;i,j;0)}{1}{2}{a}{b,\frac{\dimd-1}{2}}{-\frac{\proper{q}^2}{4}}$ in \eqref{eq:A2singular} and  \eqref{eq:A2singularh}, we express the derivatives as the limit of a finite difference. Exchanging the Fourier transform with the limit, the finite difference is of the form \eqref{eq:finitedifference}.
	
	The potential is given by a discontinuous function, 
	\begin{equation}\label{eq:potentialdiscontinuous}
		V_i(\proper{r})	=	\begin{cases}
			V_i^{\proper{r}<1}(\proper{r})	&	\proper{r} < 1 \\
			0	& \proper{r} >1
		\end{cases}
		\,\text{.}
	\end{equation}
	The  potentials $V_i^{\proper{r}<1}$ are conveniently expressed in terms of the following functions, valid for $\proper{r}<1$ and non-negative integers $n$ and $m$:
	\begin{equation}\begin{aligned}
			\mathcal{V}_{nm}(\proper{r};\zeta)	&=	\frac{2^{2-\dimd}	\pi^{\frac{1-\dimd}{2}}}{\Gamma\left(\frac{\dimd-1}{2}\right)}	\int_0^\infty	\dd{\proper{q}}	\,	\proper{q}^{\dimd-2}	\HypergeometricZeroFOne{\frac{\dimd-1}{2}}{-\frac{\proper{q}^2\proper{r}^2}{4}}	\HypergeometricPFQtilde{1}{2}{n}{\frac{\dimd-1}{4}+m,\frac{\nu(\zeta)}{2}}{-\frac{\proper{q}^2}{4}}
			\\&=
			\begin{aligned}[t]&
				\pi^{\frac{1-\dimd}{2}}\frac{\Gamma\left(\frac{\dimd-1}{2}-n\right)}{\Gamma(n)}	\frac{\Gamma\left(\frac{\dimd-1}{4}+m-\frac{\nu(\zeta)}{2}\right)\Gamma\left(\frac{\dimd-1}{4}+m+\frac{\nu(\zeta)}{2}\right)}{\Gamma\left(\frac{\dimd-1}{4}+m-n-\frac{\nu(\zeta)}{2}\right)\Gamma\left(\frac{\dimd-1}{4}+m-n+\frac{\nu(\zeta)}{2}\right)}
				\times\\&\qquad\qquad
				\proper{r}^{1-\dimd+2n}\HypergeometricPFQtilde{2}{1}{\frac{5-\dimd}{4}-m+n,\frac{\nu(\zeta)}{2}}{\frac{3-\dimd}{2}+n}{\proper{r}^2}
				\\&
				+	\pi^{\frac{1-\dimd}{2}}	\frac{\Gamma\left(\frac{1-\dimd}{2}+n\right)}{\Gamma(n)}	\frac{\Gamma\left(\frac{\dimd-1}{4}+m-\frac{\nu(\zeta)}{2}\right)\Gamma\left(\frac{\dimd-1}{4}+m+\frac{\nu(\zeta)}{2}\right)}{\Gamma\left(\frac{1-\dimd}{4}+m-\frac{\nu(\zeta)}{2}\right)\Gamma\left(\frac{1-\dimd}{4}+m+\frac{\nu(\zeta)}{2}\right)}
				\times\\&\qquad\qquad
				\HypergeometricPFQtilde{2}{1}{\frac{\dimd+3}{4}-m,\frac{\nu(\zeta)}{2}}{\frac{\dimd+1}{2}-n}{\proper{r}^2}
				\,\text{.}	
			\end{aligned}
	\end{aligned}\end{equation}
	In this expression, we defined $\HypergeometricPFQtilde{2}{1}{a,b}{c}{z}= \HypergeometricPFQ{2}{1}{a-b,a+b}{c}{z}$ in terms of a hypergeometric function.
	It is now straightforward to express the  $V_i^{\proper{r}<1}$ in terms of $\mathcal{V}_{nm}$. For $V_0$, we find
	\begin{equation}
		V_0^{\proper{r}<1}(\proper{r};\zeta)
		=	
		\frac{\mu_\phi \mu_\chi}{2\zeta}	\mathcal{V}_{1,1}(\proper{r};\zeta)
		\,\text{.}
	\end{equation}
	Similar to $A_2$, the potential $V_2^{\proper{r}<1}$ splits into a regular and a singular part:
	\begin{equation}
		V_2(\proper{r};\zeta) = V_2^{\text{reg}}(\proper{r}) + V_2^{\text{sing}}(\proper{r})
		\,\text{.}
	\end{equation}
	The regular part is given by
	\begin{equation}
		\begin{aligned}
			V_2^{\text{reg}}(\proper{r};\zeta)
			=
			\begin{aligned}
				\frac{\mu_\phi \mu_\chi}{2(\zeta+\dimd)}	\Bigg[
				\frac{1}{\dimd-1}	\mathcal{V}_{1,0}(\proper{r};\zeta)
				+	\left[	\frac{\dimd-3}{\zeta}	-\frac{1}{\dimd}	\right]	\mathcal{V}_{1,1}(\proper{r},\zeta)
				+	\frac{2}{\zeta}	\mathcal{V}_{2,1}(\proper{r};\zeta)
				\Bigg]
				\,\text{,}
			\end{aligned}
	\end{aligned}\end{equation}
	while the singular part becomes
	\begin{equation}\begin{aligned}
			V_2^{\text{sing}}(\proper{r};\zeta)
			=
			\begin{cases}
				\begin{aligned}\frac{\mu_\phi \mu_\chi}{2(\zeta+\dimd)}	\frac{\dimd-2}{\dimd-1}	\mathcal{V}_{1,0}(\proper{r};\zeta-\zeta_h)
					\,\text{,}\end{aligned}&	\zeta \neq \zeta_h
				\\\\
				\begin{aligned}
					-	\frac{\mu_\phi \mu_\chi}{2}	\left[
					\frac{1}{\dimd-1}	\mathcal{V}_{1,0}(\proper{r};0)
					+	\frac{\pi^{\frac{1-\dimd}{2}}	}{2}	\Gamma\left(\frac{\dimd+1}{2}\right)
					\right]
					\,\text{,}\end{aligned}&	\zeta =\zeta_h
			\end{cases}
			\,\text{.}
	\end{aligned}\end{equation}
	The function $V_2^{\text{sing}}(\proper{r};\zeta=\zeta_h)$ reduces to the rational function
	\begin{equation}
		V_2^{\text{sing}}(\proper{r};\zeta=\zeta_h)
		=
		\frac{\mu_\phi \mu_\chi}{2}	\pi^{\frac{1-\dimd}{2}}	\Gamma\left(\frac{\dimd-1}{2}\right)	\frac{\proper{r}^{3-\dimd}
			+	\frac{1}{4}-\proper{r}^2	-	\frac{1}{4}(\dimd+\proper{r}^2-\dimd \proper{r}^2)^2}{(\dimd-1)(\proper{r}-1)^2(\proper{r}+1)^2}	
		\,\text{.}
	\end{equation}
	The total, dimensionful scattering potential for \QuadG{} can easily be expressed in terms of the $V_i$. For $\proper{r}<1$, we have
	\begin{equation}\label{eq:potentialQuadG}
		\begin{aligned} V^{\proper{r}<1}(\proper{r}) = 16\pi \GN  H^{3}\, c(\aRR,\aE)	\bigg[ &
			-	\frac{\zeta_C+\dimd}{\zeta_C-\zeta_h}	\Big(	V_2(\proper{r};\zeta_h)	-	V_2(\proper{r};\zeta_C)	\Big)
			\\&
			+	\frac{1}{\dimd(\dimd-1)}	\frac{\zeta_C}{\zeta_C-\zeta_h}	\Big( V_0(\proper{r};\zeta_h)	-	V_0(\proper{r};\zeta_C)	\Big) 
			\\&
			+	\frac{1	}{(\dimd-1)(\dimd-2)}	\Big( V_0(\proper{r};\zeta_h)	-	V_0(\proper{r};\zeta_R)	\Big) 
			\bigg]
			\,\text{,}
		\end{aligned}
	\end{equation}
	while the potential is equal to zero for $\proper{r}>1$.

	\subsection{Properties of the potential}
	Having computed the potential for \QuadG{} in \eqref{eq:potentialQuadG}, we will now consider its phenomenological properties. To this end, we will regard the potential as the source of a Newtonian force. In accordance with the classical Newtonian dynamics we define the gravitational force as $F=-\nabla V(r)$.

	This allows to straightforwardly interpret the potential in terms of potential energies, and the scaling of the force with distance.
	
	\subsubsection{The short-distance regime}
	We will first consider the regime $\proper{r}\ll 1$. In this region, spacetime can be approximated as locally flat, so that we expect curvature effects to be negligible. Hence, in this limit we should recover the behavior of the potential in Minkowski spacetime. The Yukawa-like potential of \QuadG{} is easily obtained from the Minkowski-spacetime scattering amplitude, and reads
	\begin{equation}\label{eq:Yukawa}
		V_{\text{Mink}}(r) = \GN m_\phi m_\chi \left(	-\frac{1}{r}	- \frac{1}{3} \frac{1}{r}\ee^{-r/\sqrt{\aRR}} + \frac{4}{3}	\frac{1}{r}\ee^{-r/\sqrt{\aCC}}	\right)
	\end{equation}
	For \deS{} spacetime, there are four cases to be considered:
	
	\paragraph{General relativity}
	In this case, the higher-derivative couplings $\aRR$ and $\aCC$ are both zero. Upon expanding around $\proper{r}=0$, we then find
	\begin{equation}
		V_{\text{GR}}(\proper{r})	\sim	-	\GN H^3	\frac{\mu_\phi \mu_\chi}{\proper{r}}
		\,\text{.}
	\end{equation}	
	Hence, we reproduce Newton's law, as is discussed in detail in \cite{Ferrero:2021lhd}.
	
	\paragraph{$R^2$-gravity}
	We continue with $R^2$-gravity, obtained from the limit \eqref{eq:R2limit}. In this case, the $\proper{r}\to0$ behavior reads
	\begin{equation}
		V_{R^2}(\proper{r}) \sim -	\frac{4}{3}	\frac{\zeta_R}{\zeta_R+4}	\GN H^3	\frac{\mu_\phi \mu_\chi}{\proper{r}}
		\,\text{.}
	\end{equation}
	This is in agreement with the short-distance behavior for $R^2$-gravity in flat spacetime, determined by \eqref{eq:Yukawa}.
	
	At this point, a remark about Newton's constant $\GN$ is in order. 
		The numerical value of $\GN$ can be \emph{defined} as the coefficient of the $1/r$ law, which can be obtained e.g. from a Cavendish experiment. For relativistic theories, one then determines numerical prefactors by taking the appropriate Newtonian limit, and comparing coefficients. The prime example of this is the prefactor $8\pi \GN$ on the right-hand side of the Einstein equation. Therefore, in the case of $R^2$-gravity, we can renormalize the constant $\GN$ such that we obtain exactly the Newtonian potential.\footnote{Note that the sign of (the derivative of) $V$ is connected to the direction of the corresponding force. Thus,  one can only renormalize by a positive number.}
		
		Hence, setting
		\begin{equation}
			\hat{G}_{\text{N}} = \begin{cases}	\frac{4}{3}	\frac{\zeta_R}{\zeta_R+4}	\GN	\,\text{,}& 0<\zeta_R <\infty
				\\\\  \GN & \text{otherwise}
			\end{cases}
			\,\text{,}
		\end{equation}
		we conclude that also $R^2$-gravity yields a Newtonian force law at distances small compared to the Hubble length.
		
		In order to distinguish $\GN$ from $\hat{\GN}$, the universal value of $\GN$ can be determined by performing a higher order scattering experiments involving more than two particles. We emphasize anyhow that this  is unimportant for our purposes, aiming at studying the modifications to the Newtonian potential.
		
		Furthermore, we note that in the limit $\alpha_R \to 0\; (\zeta_R \to \infty)$ a discontinuity arises. This discontinuity is reminiscent of the massive gravity vDVZ discontinuity \cite{vanDam:1970vg,Zakharov:1970cc}. This arises because of the different number of degrees of freedom in the two cases: the ``massive" scalar field does not decouple properly in this limit.

	\paragraph{$C^2$-gravity}
	We proceed with $C^2$-gravity, given by the limit \eqref{eq:C2limit}. Expanding around $\proper{r}=0$ gives
	\begin{equation}
		V_{C^2}(\proper{r}) \sim +\frac{1}{3}	\GN H^3	\frac{\mu_\phi \mu_\chi}{\proper{r}}
		\,\text{,}
	\end{equation}
	which is the same as what we find from \eqref{eq:Yukawa}.
	In contrast to GR and $R^2$-gravity, we find that $C^2$-gravity yields a \emph{repulsive} force-law at short distances.
	
	\paragraph{Quadratic gravity.}
	We finish our discussion of the short-distance regime by considering full \QuadG. Here we find that the potential does not possess a pole at $\proper{r}=0$, similar to what we find by evaluating \eqref{eq:Yukawa} at $r=0$. The  absence of such a short-distance singularity can be seen as the classical analogue of the perturbative renormalizability of \QuadG, c.f. \cite{Stelle:1976gc,Stelle:1977ry}.
	
	\subsubsection{The \deS{} horizon}
	The second property of the potential that we discuss is the regime $\proper{r}\sim 1$. As we have seen in \eqref{eq:potentialdiscontinuous}, the potential vanishes at distances $\proper{r}>1$. As was already observed in \cite{Ferrero:2021lhd}, this is a natural manifestation of \deS{} horizon: since particles that are separated by the horizon are not in causal contact, their scattering potential must be identically zero.
	
	In order to further study the discontinuity, we interpret $V$ as the source of a Newtonian force. We will be interested in the dimensionless quantity
	\begin{equation}
		\proper{F} = - \frac{V'(\proper{r})}{\hat{G}_{\text{N}} H^3 \mu_\phi \mu_\chi}
		\,\text{,}
	\end{equation}
	which can be seen as the Newtonian force expressed in Hubble units.
	
	It is clear that for radii $\proper{r}>1$, the force $\proper{F}$ is zero. Approaching $\proper{r} = 1$ from below, the discontinuity in $\proper{F}$ can be computed \cite{Hawking:1975vcx, Damour:1976jd,Li:2016sjq}. We find that the discontinuity lies in the interval
	\begin{equation}
		\lim_{\proper{r}\nearrow1}\proper{F}_{\zeta\to0} \leq \lim_{\proper{r}\nearrow1}\proper{F} \leq \lim_{\proper{r}\nearrow1}\proper{F}_{\text{GR}}
		\,\text{,}
	\end{equation}
	where $\proper{F}_{\text{GR}}$ is the force for General Relativity and $\proper{F}_{\zeta\to0}$ is the force of \QuadG{} in the limit $\zeta_C,\zeta_R \to 0$. These values can be computed exactly:
	\begin{equation}
		\begin{aligned}
			\lim_{\proper{r}\nearrow1}\proper{F}_{\zeta\to0}	&= -	\frac{17}{24}	-	\frac{5\Gamma\left(\frac{3}{4}-\frac{\nu(\zeta_h)}{2}\right)	\Gamma\left(\frac{3}{4}+\frac{\nu(\zeta_h)}{2}\right)}{6\sqrt{\pi}}		+	\frac{4\Gamma\left(\frac{3}{4}-\frac{\nu(-\zeta_h)}{2}\right)	\Gamma\left(\frac{3}{4}+\frac{\nu(-\zeta_h)}{2}\right)}{3\sqrt{\pi}}	
			\\&	\approx	1.647
			\,\text{;}\\
			\lim_{\proper{r}\nearrow1}\proper{F}_{\text{GR}}	&= 	\frac{5}{6}	+	\frac{5\Gamma\left(\frac{3}{4}-\frac{\nu(\zeta_h)}{2}\right)	\Gamma\left(\frac{3}{4}+\frac{\nu(\zeta_h)}{2}\right)}{3\sqrt{\pi}}	
			\\&	\approx	3.443
			\,\text{.}
		\end{aligned}
	\end{equation}
	We notice that both values are positive. Hence, at the \deS{} horizon, the net gravitational force is repulsive, in contrast to classical Newtonian gravity. We interpret this as an effect of the positive curvature: the expansion of the universe manifests itself as a repulsive effective force.
	
	\subsubsection{MOND from quadratic gravity}
	We now study $\proper{F}$ in the intermediate regime $0 < \proper{r} <1$. We will pay special attention to the comparison to the classical Newtonian force $\proper{F}_{\text{cl}} = -\proper{r}^{-2}$. Any deviation from this potential can be seen as an instance of Modified Newtonian Dynamics (MOND). In recent years, phenomenological models of MOND have been constructed to explain rotation curves of galaxies, which do not correspond to classical Newtonian gravity 						
	\cite{Mannheim:2010xw,Milgrom:2011kx, Famaey:2011kh, 2016PhRvL.117t1101M, OBRIEN2018433, deAlmeida:2018kwq, Mannheim:2021mhj,MANNHEIM2006340,Brouwer:2021nsr}. MOND is therefore an alternative to Dark Matter (DM) models, which explain the discrepancy of galactic rotation curves by the existence of non-luminous cold matter.
	
	In order to reproduce a DM-like scenario using MOND, one mimics the additional mass density by modifying the potential such that one obtains an additional attractive force. In this section, we will consider whether such a modification can arise from $\proper{F}$ by an appropriate choice of $\zeta_C$ and $\zeta_R$. Again, we distinguish four different regimes.
	
	\begin{figure}\centering
		\begin{subfigure}{.5\textwidth}\centering
			\includegraphics[width=\textwidth]{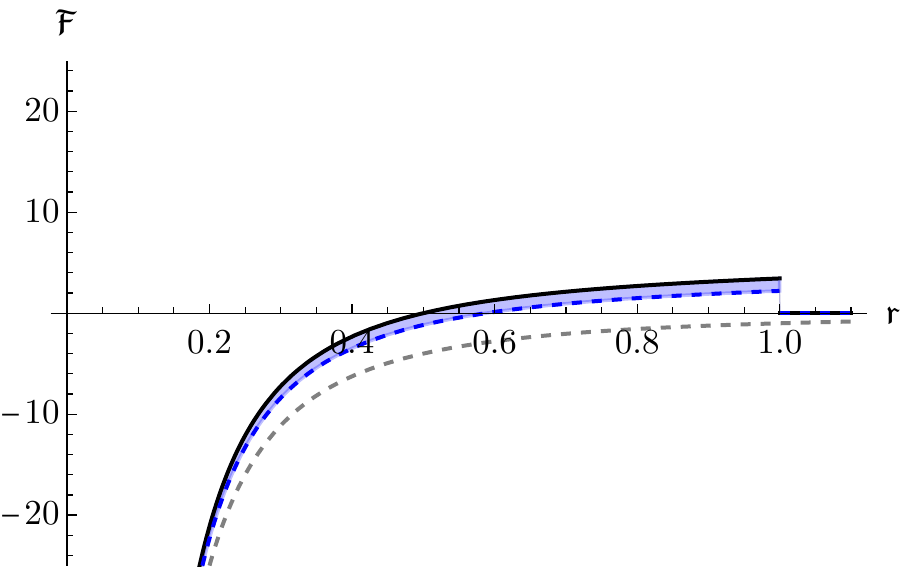}
			\caption{}
		\end{subfigure}%
		\begin{subfigure}{.5\textwidth}\centering
			\includegraphics[width=\textwidth]{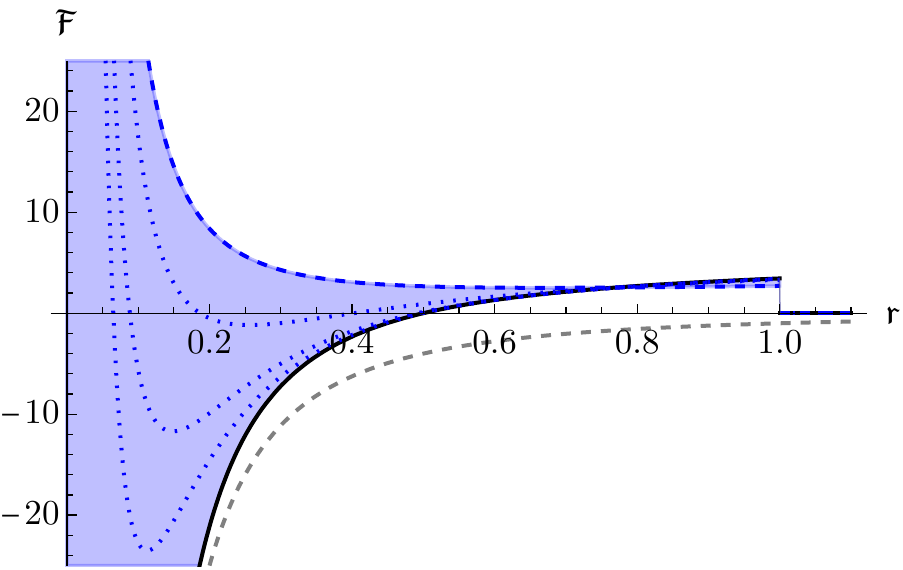}
			\caption{}
		\end{subfigure}%
		\caption{Dimensionless forces $\proper{F}$ in $R^2$- and $C^2$-gravity. In both panels, the black curve denotes $\proper{F}_{\text{GR}}$, while the dashed gray curve shows the classical Newtonian force $\proper{F}_{\text{cl}} = -\proper{r}^{-2}$ for reference. Left panel: $\proper{F}$ in $R^2$-gravity. Scanning over $\zeta_R$, the curves $\proper{F}_{\zeta_R}$ fill the shaded blue region, bounded by $\proper{F}_{\text{GR}}$ and $\proper{F}_{\zeta_R\to0}$ (dashed blue curve). Right panel: $\proper{F}$ in $C^2$-gravity. Scannosning over $\zeta_C$ fills the shaded blue region, bounded by $\proper{F}_{\text{GR}}$ and $\proper{F}_{\zeta_C\to0}$ (dashed blue curve). For typical finite values of $\zeta_C$, $\proper{F}$ is drawn using dotted blue curves. Both for $R^2$- and $C^2$-gravity, the force is strictly larger than $\proper{F}_{\text{cl}}$, indicating that a dark matter MOND-like scenario is excluded.}
		\label{fig:R2C2force}
	\end{figure}
	\paragraph{General relativity} We begin with GR. The force $\proper{F}_{\text{GR}}$ is shown by the black curve in \autoref{fig:R2C2force}. Comparing $\proper{F}_{\text{GR}}$ to the classical Newtonian force $\proper{F}_{\text{cl}}= -\proper{r}^{-2}$ (the dashed gray curve in \autoref{fig:R2C2force}), we observe that the two coincide for small $\proper{r}$, as discussed previously. Also the discontinuity is clearly visible. Furthermore, we observe that $\proper{F}_{\text{GR}}$ is strictly larger than $\proper{F}_{\text{class}}$. Hence, the spacetime curvature causes an additional expanding force, coinciding with the expectation that \deS{} spacetime models an expanding universe.
	
	\paragraph{$R^2$-gravity} Turning on the coupling $\aRR$, we find that the addition of an $R^2$-interaction gives a modification to General Relativity. Letting $\zeta_R$ run from $0$ to $\infty$, we find that the curves $\proper{F}$ fill the shaded blue region in the left panel of \autoref{fig:R2C2force}. Here the dashed blue line denotes the limiting case $\proper{F}_{\zeta_R\to 0}$. We find that the $R^2$ modification causes $\proper{F}$ to be lower than $\proper{F}_{\text{GR}}$. Hence, the $R^2$-interaction effectively causes an attractive force on top of the force due to GR. However, the modification is nowhere strong enough to give rise to DM-like MOND.
	
	\paragraph{$C^2$-gravity} In the right panel of \autoref{fig:R2C2force}, the force arising from $C^2$-gravity is shown. Similar to $R^2$-gravity, varying $\zeta_C$ traces out a region bounded by $\proper{F}_{\text{GR}}$ and $\proper{F}_{\zeta_C\to0}$. In contrast to $R^2$-gravity, $\proper{F}$ is always positive for sufficiently small $\proper{r}$, due to the $+\proper{r}^{-1}$-like behavior of the potential. In fact, below $\proper{r}\approx 0.770$, $\proper{F}$ is strictly larger than $\proper{F}_{\text{GR}}$. Therefore, in this regime the effect of the $C^2$-interaction can be interpreted as an additional repulsive force on top of General Relativity. Thus, $C^2$-gravity cannot be matched to a MOND-potential suitable to explain DM.
	
\begin{figure}\centering
		\begin{subfigure}{.5\textwidth}\centering
			\includegraphics[width=\textwidth]{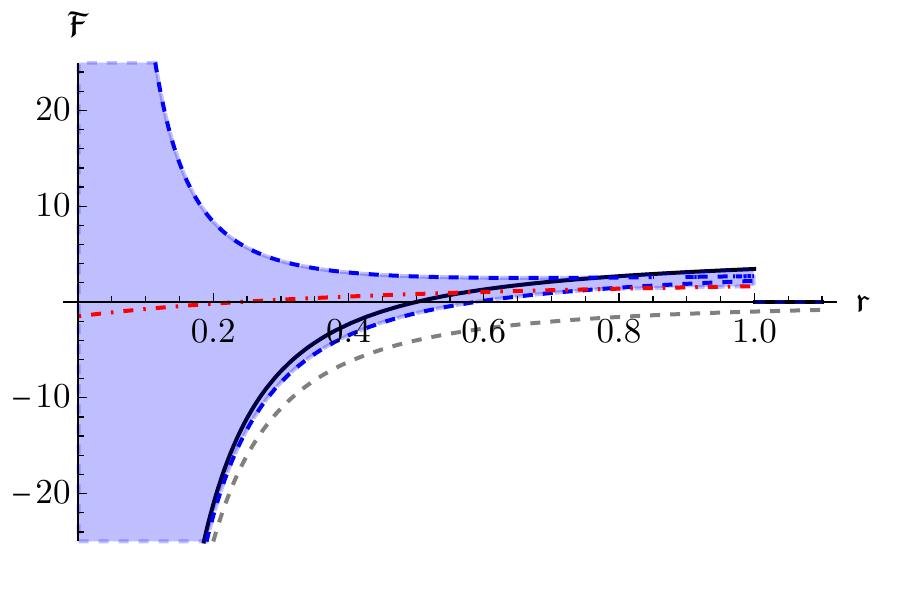}
			\caption{}
		\end{subfigure}%
		\begin{subfigure}{.5\textwidth}\centering
			\includegraphics[width=\textwidth]{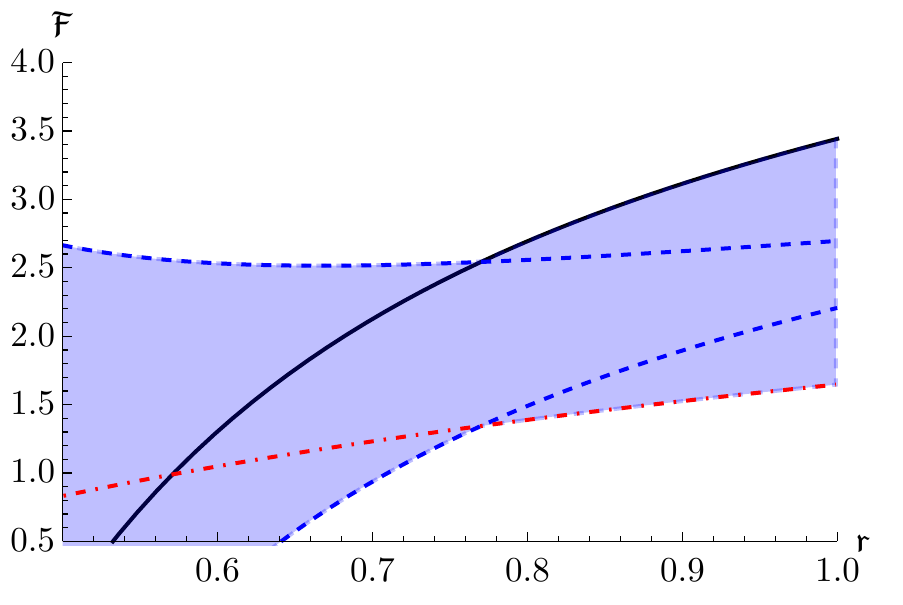}
			\caption{}
		\end{subfigure}%
		\caption{Dimensionless forces $\proper{F}$ in \QuadG. The black curve denotes $\proper{F}_{\text{GR}}$. The dashed gray curve shows the classical Newtonian force $\proper{F}_{\text{cl}} = -\proper{r}^{-2}$. For any point in the shaded blue region, there exist values of $\zeta_C$, $\zeta_R$ such that this point is reached by $\proper{F}_{\zeta_C,\zeta_R}$. In particular, for any $\zeta_C,\zeta_R$, $\proper{F}$ reaches a finite value at $\proper{r}=0$. The shaded blue region is bounded by $\proper{F}_{\text{GR}}$, $\proper{F}_{\zeta_R\to0}$, $\proper{F}_{\zeta_C\to0}$ (dashed blue lines), and $\proper{F}_{\zeta\to0}$ (dashed-dotted red line). The right panel shows a detail of the left panel, clarifying the boundary of the shaded region near the \deS{} horizon.}
		\label{fig:QuadGforce}
	\end{figure}

	\paragraph{Quadratic gravity} Finally, we consider full \QuadG. As we have seen, since the potential of \QuadG{} is finite at $\proper{r}=0$, there will be no $\proper{r}^{-2}$-like behavior of $\proper{F}$. Instead, we find that $\proper{F} \sim -\frac{3}{2}	-	\frac{\zeta_C}{2}	+\frac{\zeta_R}{8}$ for small $\proper{r}$.  Thus, by an appropriate choice of $\zeta_C$ and $\zeta_R$, the force $\proper{F}$ can reach any finite value at $\proper{r}=0$. This includes $\proper{F}=0$, which can be related to force laws arising from spacetimes  with higher regularity \cite{Bonanno:2000ep}.   We find that $\proper{F}$ is mostly bounded by $\proper{F}_{\zeta_R\to0}$ and $\proper{F}_{\zeta_C\to0}$ (the shaded blue region in \autoref{fig:QuadGforce}). For $\proper{r} \gtrsim 0.770$, the force is bounded by $\proper{F}_{\text{GR}}$ and $\proper{F}_{\zeta\to0}$, obtained by taking the limit $\lim_{\zeta_C,\zeta_R\to 0}\proper{F}$. This is shown in detail in the right panel of \autoref{fig:QuadGforce}.
	
	Thus, we conclude that no choice of $\zeta_C$ and $\zeta_R$ gives rise to an effective force that is more attractive than the classical Newtonian force. Instead, we find in the entire parameter space a repulsive contribution to the gravitational force. Therefore, \deS{} curvature corrections to \QuadG{} cannot explain galactic rotation curves.
	
	\section{Conclusion}\label{sec:conclusion}
	In \cite{Ferrero:2021lhd}, we introduced a covariant framework to compute Green's functions and potentials in \deS{} spacetime. In this follow-up paper, we extend the formalism to include non-minimal interactions. More specifically, we considered the tree-level scattering of two scalar fields in quadratic gravity, given by $R^2$- and $C^2$-terms in the action, in addition to the Einstein-Hilbert action with cosmological constant. Apart from a kinetic term and mass term, the action for the scalar fields is given by an $R\phi\phi$ interaction.
	
	After having reviewed the derivation of the Yukawa potential in \autoref{sec:Yukawa}, in \autoref{sec:amplitudefunctional} we constructed the Green's functional for this scattering process. In addition to be fully covariant, we showed that the on-shell Green's functional is explicitly gauge-independent. We organized the propagator and vertices in such a way that an effective mass-pole structure for rank-0 and rank-2 vertex tensors could be identified. Novel in this setting are corrections to the poles due to the \deS{} curvature, parameterized by the parameters $\zeta_C$, $\zeta_R$ and $\zeta_h$. We observed that in the case of conformally coupled scalars, only the rank-2 vertex tensor $T_{\mu\nu}$ contributes.
	
	In order to extract the explicit expression of the tree-level amputated Green's function, we considered the adiabatic limit of the Green's functional in \autoref{sec:adiabatic-amplitude}. This limit is implemented as an expansion in a large mass of the scalar fields compared to the Hubble constant, i.e., as an expansion in $\mu_\phi^{-1}$ and $\mu_\chi^{-1}$. This expansion allows to use spatial momentum techniques to evaluate the Green's functional.
	We performed this analysis for any mass parameter $\zeta$ of the graviton propagator, necessary to cover quadratic gravity. It turns out that the mass parameter $\zeta_h$, which represents the usual value of the mass parameter associated to the massless graviton in a de Sitter spacetime \cite{Garidi:2003ys, Pejhan:2018ofn, Joung:2006gj, Joung:2007je}, is special, in the sense that the amplitude for this mass parameter is not continuously connected to the amplitude corresponding to different masses.
	
	Analogous to what has been found in the GR case \cite{Ferrero:2021lhd}, the amplitude of quadratic gravity presents an oscillating behavior as a function of the proper transferred momentum $\proper{q}$, giving rise to an amplitude vanishing for periodic discrete values. This feature is attributed to  the presence of the horizon in a de Sitter spacetime: the particles are bounded to interact within an Hubble volume. Furthermore, the amplitude is finite for $\proper{q} \to 0$, similar to the behavior of a mass term in a flat-spacetime massive propagator. In particular, this holds too for $\zeta=\zeta_h$, corresponding to a regularization of the amplitude due to a nonzero $H$. This is in contrast to the flat-spacetime case, where a massless graviton propagator gives rise to a divergent amplitude in the limit $q\to0$.
	
	We then took in \autoref{sec:potential} the Fourier transform of the amplitude to obtain the scattering potential. For small proper separations $\proper{r}$, we find a potential that is in agreement with the flat-spacetime Yukawa potentials. Generalizing the result in \cite{Ferrero:2021lhd}, we found that the potential at super-Hubble distances is exactly zero: there is not causal interaction between particles separated by the \deS{} horizon. At the horizon $\proper{r}=1$, we computed the $\zeta$-dependent discontinuity in the potential.
	
	Interpreting the scattering potential as the source of a Newtonian force, we investigated whether the modified potentials in quadratic gravity could give rise to Modified Newtonian Dynamics corresponding to dark-matter like rotation curves. However, we report that the \deS{} curvature gives rise to an effective repulsive force that cannot be matched to a dark-matter like scenario.
	
Along this treatment, in order to obtain the classical nonrelativistic potential, the choice of the boundary conditions is not important. However, this choice should be resolved in order to construct causal and relativistic observables.
	
	This program for computing \deS{} spacetime Green's function can be investigated further in a number of ways. First, it is interesting to extract the amplitude and potential for conformally coupled scalar fields. An initial investigation of this system shows that this gives rise to a modification of the differential equations determining the amplitude. Currently it is unknown whether this system of equations can be solved.
	
	Moreover, we can investigate the Green's function by extending adiabatic expansion.  Expanding up to second order in the $\mu_\phi^{-1}$ and $\mu_\chi^{-1}$ would furnish information about the dependence of the amplitude on the scattering angle, which is usually encoded in the post-Newtonian (non-relativistic) expansion. Secondly, in order to take into account off-shell quantum corrections, it would be instructive to consider loop Feynman diagrams. An efficient way of organizing these quantum effects is by using form factors \cite{Knorr:2022dsx}.
	
	Further investigation of the discontinuity may also shed more light on the properties of the \deS{} horizon. In the present work, the discontinuity arises as a result of the Fourier transform. Here, special care has to be taken in integrating over the momentum $\proper{q}$ in the presence of the expansion in $\mu_\phi$ and $\mu_\chi$. A computation of the potential in a system where the adiabatic expansion is not needed, e.g., for conformally coupled scalars, could give more insight into the behavior of the potential around $\proper{r}=1$. 
	
	In addition, this may also yield information regarding the thermodynamic properties of the \deS{} horizon. A potential with curvature-induced modifications may accommodate additional particle states, which may be given an interpretation in terms of particle production. This can then be compared to the \deS{} temperature of the horizon.
	
	Finally, this covariant construction of Green's functionals can be extended to other curved backgrounds that also admit an adiabatic expansion, such as FLRW spacetimes. In particular, can be applied in the slow-roll inflationary spacetimes. A comparison to $n$-point functions imprinted in the Cosmic Microwave Background would allow to make contact with astrophysical observations.

	\section*{Acknowledgements}
	The authors like to thank Markus Fr\"ob and Martin Reuter for interesting discussions and helpful comments on the manuscript and the referee for her/his thoughtful comments and and careful review, which helped improve the manuscript.
	
	\appendix
	
	\section{Conventions}\label{app:conventions}
	In this Appendix, we collect some basic facts about \deS{} spacetime. The $\dimd$-dimensional \deS{} spacetime is uniquely characterized as the maximally symmetric Lorentzian manifold with constant positive scalar curvature. The Ricci curvature tensor is given by
	\begin{equation}
		R_{\mu\nu} = \frac{R}{\dimd} g_{\mu\nu}
		\,\text{,}
	\end{equation}
	where $R>0$ is the constant scalar curvature. The Weyl tensor vanishes. It is convenient to introduce the Hubble constant $H>0$:
	\begin{equation}
		R = \dimd(\dimd-1)H^2
		\,\text{.}
	\end{equation}
	\DeS{} spacetime is conformally flat, which means that it is conformal to the Minkowski metric:
	\begin{equation}\label{eq:conformalcoordinates}
		\dd{s^2} = \frac{1}{(H\eta)^2} \left( - \dd{\eta^2} + \dd{x_1^2} + \cdots + \dd{x_{\dimd-1}^2}	\right)
		\,\text{.}
	\end{equation}
	In our conventions, $\eta$ runs from $-\infty $ to $0$, corresponding to past and future infinity, respectively.
	
	\section{The adiabatic expansion}\label{app:adiabatic}
	Here we compile several facts about the adiabatic expansion in \deS{} spacetime that were previously derived in \cite{Ferrero:2021lhd}. In general, adiabaticity assumes that the curvature of spacetime is much smaller than the scales associated to the processes taking place in it \cite{Agullo:2014ica,Moreno-Pulido:2022phq, Junker:2001gx, Lueders:1990np, Olbermann:2007gn, Fulling1979RemarksOP, Fulling:1974zr, Fulling:1989nb, Parker:1968mv, Parker:1969au, Parker:1971pt, Parker:1974qw, Parker:2009uva, Birrell_adiabatic, Bunch_adiabatic, Bunch:1978aq, Haro:2008zz, Winitzki:2005rw, Zeldovich1,Zeldovich2}. For quantum field theory in curved spacetime, the adiabatic expansion has the advantage that it does not rely on a strict definition of asymptotic states \cite{Marolf:2012kh}. This is of particular interest in our work.
	
	In the case of scalar particle scattering in \deS{} spacetime, the adiabatic expansion is implemented concretely by performing an expansion in $\mu^{-1} = H/m$, where $m$ is the mass of the scalar field and $H$ the Hubble parameter. Essentially, in this expansion the Compton frequency of the scalar fields is assumed to be much larger than the expansion rate of \deS{} spacetime.

	This expansion is applied to the wave functions of the scalar fields. In the conformal coordinates\footnote{Note that different foliations, which correspond to different boundary conditions, lead to different quantization schemes, see for instance \cite{deBoer:2003vf, Colosi:2010fb, Cheung:2016iub, Park:2019amz,Saharian:2021zjc}. Using different coordinate systems, the labeling of the quantum numbers and hence the unitary representation is different: these are related through a non-trivial transformation between the relative complete basis. In fact, the Green's function  and the potential are not affected by the choice of the foliation.} \eqref{eq:conformalcoordinates} solutions to the Klein-Gordon equation are labeled by the comoving momentum $\vec{p}$ and are given by \cite{Bunch,Mukhanov:2007zz}:
	\begin{align}\label{eq:hankelwaves}
		h_{\vec p,\mu}(\eta,\vec x) = \eta^{\frac{\dimd-1}{2}}	H^{(1)}_{\imath\sqrt{\mu^2-\left(\frac{\dimd-1}{2}\right)^2}}(-p\eta) \ee^{\imath \vec p \cdot \vec x}
		\,\text{.}
	\end{align}
	Here, $H^{(1)}_\nu(z)$ is the Hankel function of the first kind \cite{olver10}. 
	
	This particular choice for parameterizing the solutions to the Klein-Gordon equation leads to the Bunch-Davies vacuum \cite{Bunch, Birrell, Mukhanov:2007zz}. This generalizes the Minkowski vacuum to \deS{} spacetime by two features: it is the unique choice of solutions such that firstly the coefficients of the corresponding creation and annihilation operators are complex conjugates, and secondly such that an expansion around $\mu = \infty$ exists. 
	
	In order to expand \eqref{eq:hankelwaves} in large $\mu$, we compute the expansion of $H^{(1)}_{\imath \mu}(z)$ for large $\mu$. We observe that since the Hankel functions are  solutions to Bessel's equation
	\begin{equation}\label{eq:besselsequation}
		z^2 H''(z)	+	z H'(z)	+	(z^2 + \mu^2) H(z) = 0
		\,\text{,}
	\end{equation}
	one can expand  the equation by making the ansatz that the derivative of $H$ can be expressed in terms of $H$ itself, i.e.,
	\begin{equation}
		H'(z)	=	f(z) H(z)
		\,\text{.}
	\end{equation}
	This turns Bessel's equation \eqref{eq:besselsequation} to the nonlinear equation
	\begin{equation}\label{eq:nonlinearbessel}
		z^2  f'(z)	+	z^2  f(z)^2	+	z  f(z)	+	z^2	+	\mu^2	=	0
		\,\text{.}
	\end{equation}
	Seeking a solution that has at most a simple pole in $\mu^{-1}$, the following ansatz for $f$ is in order:
	\begin{equation}
		f(z) = \mu \sum_{n\geq 0}	 f_{n}(z) \mu^{-n}
		\,\text{.}
	\end{equation}
	Plugging this back into \eqref{eq:nonlinearbessel}, we are now able to solve the differential equation order by order in $\mu$. The first few equations are explicitly:
	\begin{equation}\begin{aligned}
			0&=	1	+	z^2	 f_0^2
			\,\text{,}&\qquad&
			0=	 f_0	+	2z	 f_0	 f_1	+	z  f_0'
			\,\text{,}\\
			0&=	z^2	+	z^2  f_1^2	+	2z^2	 f_0  f_2	+	 z^2  f_1'
			\,\text{,}&\qquad&
			0=	 f_2	+	2z  f_1  f_2	+	2z	 f_0  f_3+z  f_2'
			\,\text{.}
	\end{aligned}\end{equation}
	Solving these equations furnish the expansion in terms of $\mu^{-1}$:
	\begin{equation}\label{eq:hankelexpansion}
		f_\pm(z)
		=
		\pm	\frac{\imath \mu}{z}
		\pm	\frac{\imath z}{2\mu}
		-	\frac{z}{2\mu^2}
		-	\pm	\frac{\imath z(4+z^2)}{8\mu^3}
		+	\order{\mu^{-4}}
		\,\text{.}
	\end{equation}
	We now have to show that this solution of \eqref{eq:nonlinearbessel} indeed gives an expansion of the Hankel functions. To this end, we expand $H^{(1)}_{\imath \mu}(z)$ first around $z=0$, and subsequently around $\mu = \infty$. This gives
	\begin{equation}
		H^{(1)}_{\imath \mu}(z)
		=
		\frac{\imath \mu}{z}	+	\order{z,\mu^{-1}}
		\,\text{.}
	\end{equation}
	We note that this expansion matches exactly the first term of $ f_+$ hence $ f_+ $ represents indeed the expansion of $H^{(1)}_{\imath \mu}$. 
	
	With this expansion at hand, it is straightforward to compute the action of derivative operators acting on the mode functions \eqref{eq:hankelwaves}. In particular, we have time derivative
	\begin{equation}\label{eq:adiabaticwavefunction}
		\partial_\eta h_{\vec{p},\mu} = \left[	\frac{\imath \mu}{\eta} + \frac{\dimd-1}{\eta} + \imath \left( \frac{1}{2}p^2\eta^2 - \frac{(\dimd-1)^2}{8\eta}\right) \frac{1}{\mu} + \order{\mu^{-2}}	\right] h_{\vec{p},\mu}
		\,\text{.}
	\end{equation}
	This formula lies at the basis of the expansion of, e.g., the action of the graviton propagator $\mathcal{G}(\square)$ on the vertex tensors $V$ and $T_{\mu\nu}$ used  in this work.
	
	\section{Computational details of the Green's function in the adiabatic expansion}\label{app:details}
	
	In this Appendix, we assemble several details regarding the calculation of the scattering amplitude.
	First, we discuss the existence of the expansion of the vertex tensors $V$ and $T_{\mu\nu}$ in the adiabatic limit $\mu^{-1} \to 0$. Second, we will provide additional details regarding the adiabatic expansion of the propagator.
	
	\subsection{Adiabatic expansion of the vertices}
	Having studied the adiabatic limit of a single scalar field in \autoref{app:adiabatic}, we now consider the adiabatic expansion of the vertex tensors $V$ and $T_{\mu\nu}$. Computing the action of $\square$ on $\chi_1 \chi_2$ using \eqref{eq:adiabaticwavefunction}, we find that the vertex $V[\chi_1,\chi_2]$ expands to
	\begin{equation}
		V[\chi_1,\chi_2]
		=
		\left[
		H^2 \mu_\chi^2	+	\order{\mu_\chi^0}
		\right]	\chi_1 \chi_2
		\,\text{.}
	\end{equation}
	Thus, the terms $\mathcal{A}_0(\zeta)$ will each contribute to leading order $\mu_\chi^2$ to the Green's function.
	For the tensor vertex $T_{\mu\nu}$, we compute the components
	\begin{equation}\label{eq:T2expansion}
		\begin{aligned}
			T_{00}[\chi_1,\chi_2]	&=	\left[
			\frac{\dimd+1}{\dimd}	\frac{1}{\eta^2}	\, \mu_\chi^2
			+	\order{\mu_\chi^{0}}
			\right]	\chi_1	\chi_2
			\,\text{,}
			\\
			T_{0i}[\chi_1,\chi_2]	&=	\left[
			\frac{1}{2\eta}	(k_{1,i}+ k_{2,i})	\,\mu_\chi
			+	\order{\mu_\chi^{0}}
			\right]	\chi_1	\chi_2
			\,\text{,}
			\\
			T_{ij}[\chi_1,\chi_2]	&=	\left[\frac{1}{d \eta^2} \delta_{ij} \,\mu_\chi^2
			+	\order{\mu_\chi^{0}}
			\right]	\chi_1	\chi_2
			\,\text{.}
	\end{aligned}\end{equation}
	Note that the (00)-component and the $(ij)$-components are of order $\mu_\chi^2$. Therefore, only these will contribute to the leading order in the adiabatic expansion of $\mathcal{A}_2(\zeta)$.
	
	\subsection{Expansion of the propagator}
	We proceed by showing that the action of the propagator on the scalar fields also has a well-behaved adiabatic expansion. We will show this explicitly for the scalar vertex $V$; for the tensor vertex $T_{\mu\nu}$, the procedure is completely analogous.
	
	First, we show that $f(\square) \phi_1\phi_2$ has an expansion in $\mu_\phi^{-1}$ for any analytic function $f$. Since $f$ has a Taylor series expansion, it suffices to show that the expansion exists for $\square^n \phi_1 \phi_2$.
	This is shown by an inductive argument. First, we note that for $n=0$, we have trivially
	\begin{equation}\label{eq:muexpansionzeroderivatives}
		\square^n \phi_1 \phi_2
		=
		\phi_1 \phi_2
		\equiv
		\left[	\alpha_0(\eta)	+	\order{\mu_\phi^{-1}}	\right]	\phi_1 \phi_2
		\,\text{,}
	\end{equation}
	namely for $\alpha_0(\eta) = 1$.
	We now make the inductive assumption that \eqref{eq:muexpansionzeroderivatives} holds for $n\geq 0$. Then acting with one more d'Alembertian gives
	\begin{equation}\label{eq:muexpansioninduction}
		\square^{n+1}	\phi_1 \phi_2
		=
		\square \left(\alpha_n(\eta)	\phi_1\phi_2\right)
		=
		\left[
		\left(q^2	\eta^2 \alpha_n
		+	\dimd \eta \alpha_n'
		+	\eta^2	\alpha_n''\right)	H^2
		+	\order{\mu_\phi^{-1}}
		\right]	\phi_1\phi_2
		\,\text{.}
	\end{equation}
	Here, the prime denotes a derivative with respect to $\eta$.
	By induction, it follows that $\square^n\phi_1\phi_2$ is $\order{\mu_\phi^{0}}\phi_1\phi_2$ for any $n$. Therefore, for any function $f$ that admits an analytic expansion, $f(\square)\phi_1\phi_2$ can be expanded to  zeroth order in $\mu_\phi^{-1}$.
	
	Analogously, one proves  that $f(\square)T_{\mu\nu}$ has a well-defined expansion, by  taking care of the tensor structure of $T_{\mu\nu}$. Here, the leading order is $\order{\mu_\phi^2}$, cf. \eqref{eq:T2expansion}. 
	
	We note from \eqref{eq:muexpansioninduction} that $\square^{n}\phi_1 \phi_2$ also contains $n$ powers of $q^2 \eta^2 = \proper{q}^2$. Thus, we conclude that the expansion of $f(\square)\phi_1 \phi_2$ is automatically analytic in $\proper{q}$.
	
	In particular, this is true for the propagator. Representing $(\square+ z)^{-1}$ as a geometric series,
	\begin{equation}
		(\square+z)^{-1} = \frac{1}{z} \sum_{n=0}^\infty \left(\frac{-\square}{z}\right)^n
		\,\text{,}
	\end{equation}
	we see that for $z\neq 0$ the propagator has an analytic expansion. Thus, we require the adiabatic expansion of the propagator in $\mu_\phi^{-1}$ to be analytic in $\proper{q}$.
	
	As a consequence, the solution to the differential equation \eqref{eq:G0equation} is fixed. Since this is a second-order inhomogeneous differential equation, one would expect the solution to be formed an inhomogeneous solution, accompanied by a two-dimensional homogeneous solution space. However, since the homogeneous solutions are generally not analytic, we will discard these in this paper.

	\bibliographystyle{JHEP}
	\bibliography{references.bib}
\end{document}